\documentclass[longauth]{aa}
\usepackage{natbib} 
\bibpunct{(}{)}{;}{a}{}{,}
\usepackage[switch]{lineno}
\usepackage{graphicx}
\usepackage{txfonts}
\pdfmapfile{+txfonts.map}
\usepackage{url}
\usepackage{booktabs,caption}
\usepackage{soul}
\usepackage{color} 

\usepackage[colorlinks=true,urlcolor=blue,citecolor=blue,linkcolor=blue,bookmarks=true]{hyperref}

\begin{document} 
   \title{Statistics of VHE $\gamma$-Rays in Temporal Association with Radio Giant Pulses from the Crab Pulsar}
   \author{MAGIC Collaboration:
M.~L.~Ahnen\inst{1} \and
S.~Ansoldi\inst{2,}\inst{3} \and
L.~A.~Antonelli\inst{4} \and
C.~Arcaro\inst{5} \and
A.~Babi\'c\inst{6} \and
B.~Banerjee\inst{7} \and
P.~Bangale\inst{8} \and
U.~Barres de Almeida\inst{8,}\inst{9} \and
J.~A.~Barrio\inst{10} \and
J.~Becerra Gonz\'alez\inst{11}
% ...
}

\institute { ETH Zurich, CH-8093 Zurich, Switzerland
\and Universit\`a di Udine, and INFN Trieste, I-33100 Udine, Italy
\and Japanese MAGIC Consortium: ICRR, The University of Tokyo, 277-8582 Chiba, Japan; Department of Physics, Kyoto University, 606-8502 Kyoto, Japan; Tokai University, 259-1292 Kanagawa,
Japan
\and National Institute for Astrophysics (INAF), I-00136 Rome, Italy
\and Universit\`a di Padova and INFN, I-35131 Padova, Italy
\and Croatian MAGIC Consortium: University of Rijeka, Department of Physics, 51000 Rijeka, University of Split - FESB, 21000 Split,  University of Zagreb - FER, 10000 Zagreb, University of Osijek, 31000 Osijek and Rudjer Boskovic Institute, 10000 Zagreb, Croatia.
\and Saha Institute of Nuclear Physics, HBNI, 1/AF Bidhannagar, Salt Lake, Sector-1, Kolkata 700064, India
\and Max-Planck-Institut f\"ur Physik, D-80805 M\"unchen, Germany
\and now at Centro Brasileiro de Pesquisas Físicas (CBPF), 22290-180 URCA, Rio de Janeiro (RJ), Brasil
\and Universidad Complutense, E-28040 Madrid, Spain
\and Inst. de Astrof\'isica de Canarias, E-38200 La Laguna, and Universidad de La Laguna, Dpto. Astrof\'isica, E-38206 La Laguna, Tenerife, Spain
% ...
}

      \date{Received ; accepted }
  \abstract 
   {}
{The aim of this study is to search for evidence of a common emission engine between radio giant pulses (GPs) and very-high-energy (VHE, E>100\,GeV) $\gamma$-rays from the Crab pulsar.}
{16 hours of simultaneous observations of the Crab pulsar at 1.4\,GHz with the Effelsberg radio telescope and the Westerbork Synthesis Radio Telescope (WSRT), and at energies above 60\,GeV with the Major Atmospheric Gamma-ray Imaging Cherenkov (MAGIC) telescopes were performed. We searched for a statistical correlation between the radio and VHE $\gamma$-ray emission with search windows of different lengths and different time lags to the arrival times of a radio GP. A dedicated search for an enhancement in the number of VHE $\gamma$-rays correlated with the occurrence of radio GPs was carried out separately for the P1 and P2 phase ranges respectively.}
{99444 radio GPs have been detected in the radio data sample. We find no significant correlation between the GPs and VHE photons in any of the search windows. Depending on phase cuts and the chosen search windows we find upper limits at 95\% confidence level on an increase in VHE $\gamma$-ray events correlated with radio GPs between 7\% and 61\% of the average Crab pulsar VHE  flux for the P1 and P2 phase ranges respectively. This puts upper limits on the flux increase during a radio GP of 12\% to 2900\% (depending on search window duration and phase cuts) of the pulsed VHE flux. 
This is the most stringent upper limit on a correlation between $\gamma$-ray emission and radio GPs reported so far.}
   {}
\titlerunning{Statistics of VHE $\gamma$-Rays in Temporal Association with Radio Giant Pulses from the Crab Pulsar}
\authorrunning{M.~L.~Ahnen et al.} 
\maketitle

\renewcommand{\thefootnote}{\fnsymbol{footnote}}
\footnotetext[0]{$^{\textcolor{blue}{\star}}$ Corresponding authors:
\newline 
\href{mailto:natalia.lewandowska@mail.wvu.edu}{natalia.lewandowska@mail.wvu.edu}
\newline 
\href{mailto:tsaito@icrr.u-tokyo.ac.jp}{tsaito@icrr.u-tokyo.ac.jp}}
\renewcommand{\thefootnote}{\arabic{footnote}}
\setcounter{footnote}{0}
 
%
%-------------------------------------------------------------------
\section{Introduction}\label{sec:intro}
Since the discovery of the first pulsar \citep{1968Natur.217..709H}, more than 2500 of these objects have been found \citep{2005AJ....129.1993M}. They have been observed in a large variety regarding their emission properties which led to the designation of diverse populations in the literature \citep[see review by][]{2013FrPhy...8..679H}. Some pulsars are observed only at certain wavelengths, while others can be observed throughout large parts of the electromagnetic spectrum. The Crab pulsar has been observed so far from about $10^{-8}$~eV \citep[20 MHz,][]{2013ApJ...768..136E} up to $1.5\cdot10^{12}$~eV \citep{2016A&A...585A.133A}. The approximate alignment of its pulsed emission across the electromagnetic spectrum \citep[time delays were reported by][]{2008A&A...488..271O} suggests a common engine for its broadband pulsed emission. The Crab pulsar is therefore a suitable object to test various emission theories explaining the generation of its multi-wavelength emission.\par
The average pulse profile of the Crab pulsar changes with frequency, showing up to seven different components \citep{1996ApJ...468..779M,2015ApJ...802..130H}. In the radio band, below 5~GHz, it consists of the "Main Pulse" (MP, at a rotation phase from $\sim$ -0.01 to $\sim$ 0.01), the "Low Frequency Interpulse" (LFIP, from $\sim$ 0.39 to $\sim$ 0.42 in phase), the "Precursor" (PC, from $\sim$ -0.07 to $\sim$ -0.02 in phase, only below 0.6~GHz) and the "Low Frequency Component" (LFC, from $\sim$ -0.14 to $\sim$ -0.07 in phase, only between 0.6 and 4.2~GHz). At above 5~GHz, the MP vanishes, and an additional interpulse component known as "High Frequency Interpulse" (HFIP) occurs which is shifted by about 0.02 with regard to the LFIP and located at $\sim$ 0.36 to $\sim$ 0.42 in phase. In addition, two components known as High Frequency Components (HFC1 at $\sim$ 0.53 to $\sim$ 0.67 in phase, HFC2 at $\sim$ 0.68 to $\sim$ 0.81 in phase) appear  \citep{1996ApJ...468..779M,2015ApJ...802..130H}. The names of all components and corresponding phases are summarized in Table~\ref{table:rotational_phases}. \par

\begin{table*}
\caption{Rotational phase ranges, frequency ranges of occurrence
and nomenclature of average emission components of the Crab pulsar.}
\label{table:rotational_phases}
\centering
\begin{tabular}{|c|c|c|c|c|}
\hline
 \multicolumn{3}{|c|}{Radio} & \multicolumn{2}{c|}{$\gamma$-ray} \\
\hline
  Name & Phase Range$^{1}$ & Frequency$^{1}$ & Name & Phase Range$^{2}$\\
  & [Periods] & [GHz] & & [Periods]\\
\hline
 LFC & $\sim$ -0.14 to $\sim$ -0.07 & 0.6 - 4.2 &  &  \\
 \cline{1-3}
 PC & $\sim$ -0.07 to $\sim$ -0.02 & 0.3 - 0.6 & {\bf P1} & $\sim -0.01$ to $\sim 0.02$ \\
  \cline{1-3}
 {\bf MP} & $\sim$ -0.01 to $\sim$ 0.01 & 0.3 - 4.9 &  &  \\
  \hline
   \multicolumn{3}{|c|}{}& Bridge & $\sim 0.02$ to $\sim 0.37$ \\
  \hline
 HFIP & $\sim$ 0.36 to $\sim$ 0.42 & 4.2 - 28.4 & {\bf P2}  & $\sim 0.37$ to $\sim 0.42$ \\
  \cline{1-3}
 {\bf LFIP} & $\sim$ 0.39 to $\sim$ 0.42  & 0.3 - 3.5 &  &  \\
  \hline
 HFC1 & $\sim$ 0.53 to $\sim$ 0.67 & 1.4 - 28.0 & off-pulse & $\sim 0.52$ to $\sim 0.87$ \\
  \cline{1-3}
 HFC2 & $\sim$ 0.68 to $\sim$ 0.81 & 1.4 - 28.0 &  &  \\
\hline
\end{tabular}\\
\centering $^{1}$ Radio phase ranges and frequency values are taken from \cite{2015ApJ...802..130H}.\\ 
\centering $^{2}$ $\gamma$-ray phase ranges are taken from \cite{2014A&A...565L..12A}.
\end{table*}

Extensive single pulse studies of the Crab pulsar below and above about 5~GHz show even more complex features \citep{2016ApJ...833...47H}. While MP and LFIP single pulses consist of several microsecond long bursts which can be resolved into single pulses of nanoseconds duration with continuous spectra across the observing band, HFIP single pulses consist of one burst of emission of several microseconds duration with non-uniform spectra in the form of proportionally spaced emission bands \citep{2016ApJ...833...47H}. No single pulses of nanosecond duration were detected in the case of HFIP single pulses. The observed differences therefore suggest similar emission physics for MP and LFIP single pulses and different ones for HFIP single pulses \citep{2016ApJ...833...47H}.\par
Single pulses whose flux density is more than 10 times higher than the mean are called giant pulses (GPs, \citealt{2010A&A...515A..36K}). The pulse widths of radio GPs from the Crab pulsar are in the microseconds to nanoseconds range \citep{2003Natur.422..141H} and their intensity distributions can be described by a power-law \citep{1972ApJ...175L..89A}. The shortest widths observed so far have been reported to be less than 0.4\,ns, resulting in a brightness temperature of about $10^{41}$\,K \citep{2007ApJ...670..693H}. The high brightness temperatures imply a coherent emission mechanism \citep{2009astro2010S.112H}. Strong and frequent radio GPs are observed mainly at the phase ranges of MP, LFIP and HFIP \citep{2010A&A...524A..60J,2012ASPC..466...65H}. Such a complex evolution of the average profile in radio wavelengths has never
been observed in any other pulsar so far. \par
In the $\gamma$-ray band the average pulse profile is smoother and broader than at radio frequencies \citep{kuiper_2001,abdo_2010_II,2011Sci...334...69V,2012A&A...540A..69A}. Rotation phases between -0.01 to $\sim$0.1 are often called the "P1", the ones between 0.3 and 0.5 are the "P2", and the ones between $\sim$0.1 and 0.3 are known as the "Bridge" \citep{1998ApJ...494..734F,2014A&A...565L..12A}. Note that the MP is included in the P1 range, while LFIP and HFIP are in the P2 range 
as shown in Table~\ref{table:rotational_phases}. \par
Because of the above mentioned high energy density of GPs in small volumes, a correlation between radio GPs and emission at higher energy bands can be hypothesized \citep[e.g.][]{2016JPlPh..82c6302E}. 
One  process that  could  facilitate  the  required  energy  release  on short spatial and  temporal scales is  magnetic  reconnection  in the current sheet outside the light cylinder. In this process, kinetic instabilities break the frozen-in condition of ideal magnetohydrodynamics \citep[MHD;][]{contopoulos_2010,tchekhovskoy_2013} which holds at large\footnote{larger than the kinetic length scales in the plasma.}
scales and converts magnetic energy into kinetic energy of high energy particles. Both particle-in-cell simulations \citep{spitkovsky_2006,cerutti_2012} and analytical descriptions \citep{contopoulos_1999,contopoulos_2007} of the pulsar magnetosphere confirmed the existence of current sheets, and showed the important role that the magnetic reconnection mechanism can play \citep{uzdensky_2014}. Each stochastically  occurring  reconnection  event  would  produce  radio  and  high  energy  emission  from  e.g. synchrotron emission of the energetic particles. Even if a comprehensive theoretical framework does not exist yet, the possibility of finding such a correlation between radio and $\gamma$-rays triggered different observations in the $\gamma$-ray band.\par
The Crab pulsar has been also extensively studied in the very-high-energy (VHE) $\gamma$-ray range. Imaging Air Cherenkov Telescopes (IACTs) like MAGIC (Major Atmospheric Gamma-ray Imaging Cherenkov telescopes) and VERITAS (Very Energetic Radiation Imaging Telescope Array System) revealed that the P2 component is dominant above 50~GeV up to 1.5~TeV, while the P1 component has been measured up to 600~GeV \citep{2008Sci...322.1221A,2011Sci...334...69V,2012A&A...540A..69A,2016A&A...585A.133A}. The bridge emission is significantly detected only up to $\sim$150~GeV \citep{2014A&A...565L..12A}. 
Any pulsed emission above 25~GeV cannot be explained by the conventional polar-cap pulsar models \citep{1975ApJ...196...51R,1982ApJ...252..337D,2004AdSpR..33..552B} and challenges the slot-gap scenario~\citep{2008ApJ...680.1378H}, while outer-gap models in which $\gamma$-rays are produced by curvature radiation of electrons accelerated in the magnetosphere~\citep{2008ApJ...688L..25H,2008ApJ...676..562T} are favored. \par
In the present work, we explore the association between radio GP and VHE (E~$>$ 100~GeV) $\gamma$-rays.
Separate analyses have been conducted in order to search for evidence of common emission between GPs and VHE $\gamma$-rays for each of the two $\gamma$-ray peaks P1 and P2 (and corresponding radio phases MP and LFIP respectively). From now on, adopting the notation of \citet{2014A&A...565L..12A}, we will refer to \textit{P1 GPs} and \textit{P2 GPs} to indicate GPs falling inside the VHE $\gamma$-rays phase range [-0.01 to 0.02] and [0.37 to 0.42] respectively: due to the radio frequency considered in the present work (1.4~GHz), this will translate to MP GPs and LFIP GPs. \par
Given that the origin of GPs is not known, it is certainly interesting to search for a correlation between radio GPs and VHE pulsed photons, although there is currently no theoretical approach which describes their correlation. In fact, several searches for multi-wavelength counterparts of radio GPs and optical photons were reported with 7.8 $\sigma$ \citep{2003Sci...301..493S} and 7.2 $\sigma$ \citep{2013ApJ...779L..12S} significance for MP GPs and 1.75 $\sigma$ \citep{2003Sci...301..493S} and 3.5 $\sigma$ \citep{2013ApJ...779L..12S} for LFIP GPs. This result implies the existence of an additional incoherent emission mechanism associated with radio GPs from the Crab pulsar. Similar studies were carried out in the X-ray band, finding no correlation \citep{2012ApJ...749...24B,2013ffep.confE..58M,2014efxu.conf..180M,2018PASJ...70...15H}.\par
Past searches for a correlation between radio GPs and $\gamma$-rays from the Crab pulsar provided no positive results either \citep{1974NCimB..24..153A,1995ApJ...453..433L,2011ApJ...728..110B,2012ApJ...760...64M}. The only other recent study for which data from an IACT was used was carried out by VERITAS \citep{2012ApJ...760..136A}, who searched for a correlation between radio GPs at 8.9~GHz and VHE $\gamma$-rays with energies higher than 150~GeV. With a total overlap of 11.6~h, they reported upper limits of 5 to 10 times the average Crab pulsar VHE flux on the flux measured simultaneously with P2 GPs and of 2 to 3 times the average VHE flux on time scales of about 8 seconds around P2 GPs. The present study focuses on the search for a correlation between radio GPs from the Crab pulsar and its VHE $\gamma$-ray emission. The differences with respect to the study carried out by the VERITAS Collaboration in \citet{2012ApJ...760..136A} are the following:
 \begin{enumerate}
     \item The corresponding radio data presented here were taken at a center frequency of about 1.4~GHz, whereas radio observations described in \cite{2012ApJ...760..136A} were carried out at 8.9~GHz. Based on the results by \cite{2016ApJ...833...47H} we are addressing a different population of radio GPs.
     \item The $\gamma$-ray observations reported here were carried out at energies above 60~GeV, where the P1 emission is pronounced, while in \citet{2012ApJ...760..136A} the energy threshold  was above~150 GeV, where the P1 emission is much fainter \citep{2012A&A...540A..69A}. The lower energy threshold of MAGIC (E$_{thr}\sim$~60~GeV) in comparison with the energy threshold of VERITAS (E$_{thr}\sim$~150~GeV) allows a more comprehensive analysis of the correlation between VHE $\gamma$-rays and the P1 GPs.
     \item The amount of simultaneous observations between VHE $\gamma$-rays and radio is  larger in the present study (16~h vs 11.6~h), corresponding to the currently largest sample of simultaneous VHE $\gamma$-rays and radio GP data taken with an IACT.
 \end{enumerate}

The paper is organized as follows: the observations and data analysis are described in Sec.~\ref{sec:observations}. The construction of the Monte Carlo (MC) simulations is described together with the correlation study in Sec.~\ref{sec:correlation}. The results are discussed in Sec.~\ref{sec:discussion} and a summary can be found in Sec.~\ref{sec:summary}. The appendix \ref{sec:appendix_MC_sim} carries a detailed explanation of the MC simulations developed specifically for this study.
%--------------------------------------------------------------------
\section{Observations and Data Reduction}
\label{sec:observations}
\subsection{Radio Observations}

  \begin{figure*}
   \centering 
      \includegraphics[width=0.6\textwidth ]{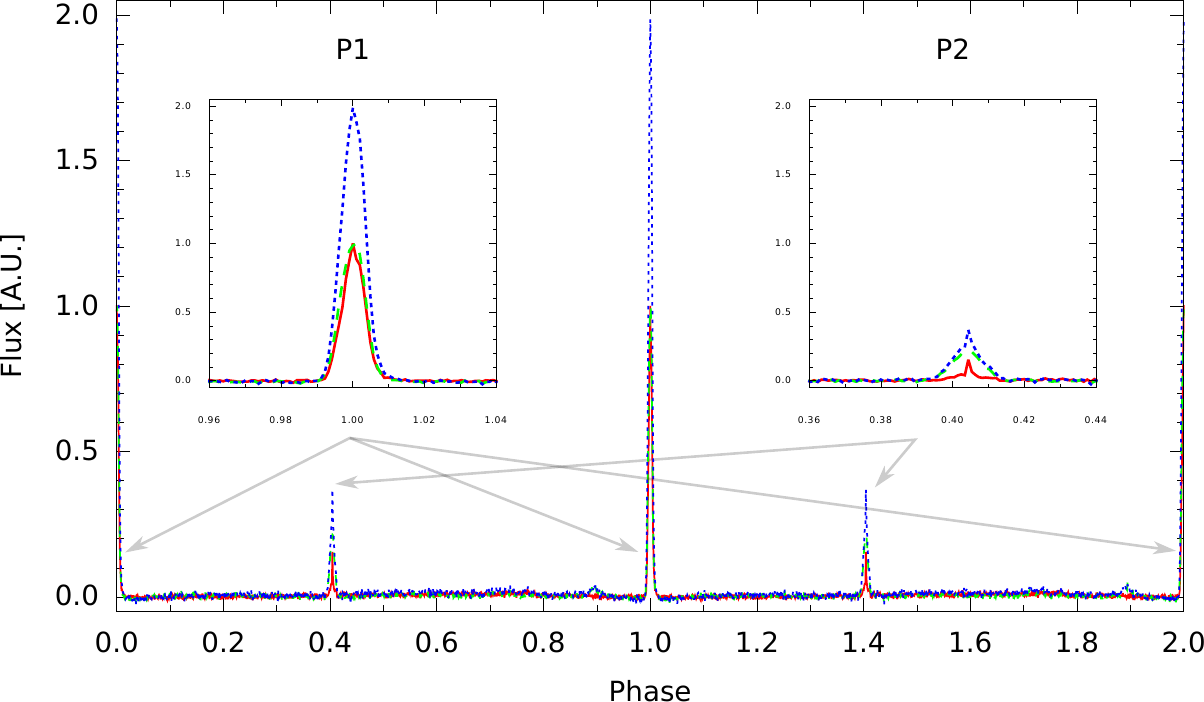}
   \caption{Phase diagram resulting from one Effelsberg observation (2017-12-07, red solid curve) and one WSRT observation (2017-12-10, green dashed curve). MP is visible near phase 0.0 and 1.0 whereas LFIP near phases 0.4 and 1.4. The blue dotted curve represents the sum of both observations.}
              \label{fig:phasediagram_radio}
    \end{figure*}
    
  \begin{figure*}
   \centering
   \includegraphics[width=0.6\textwidth]{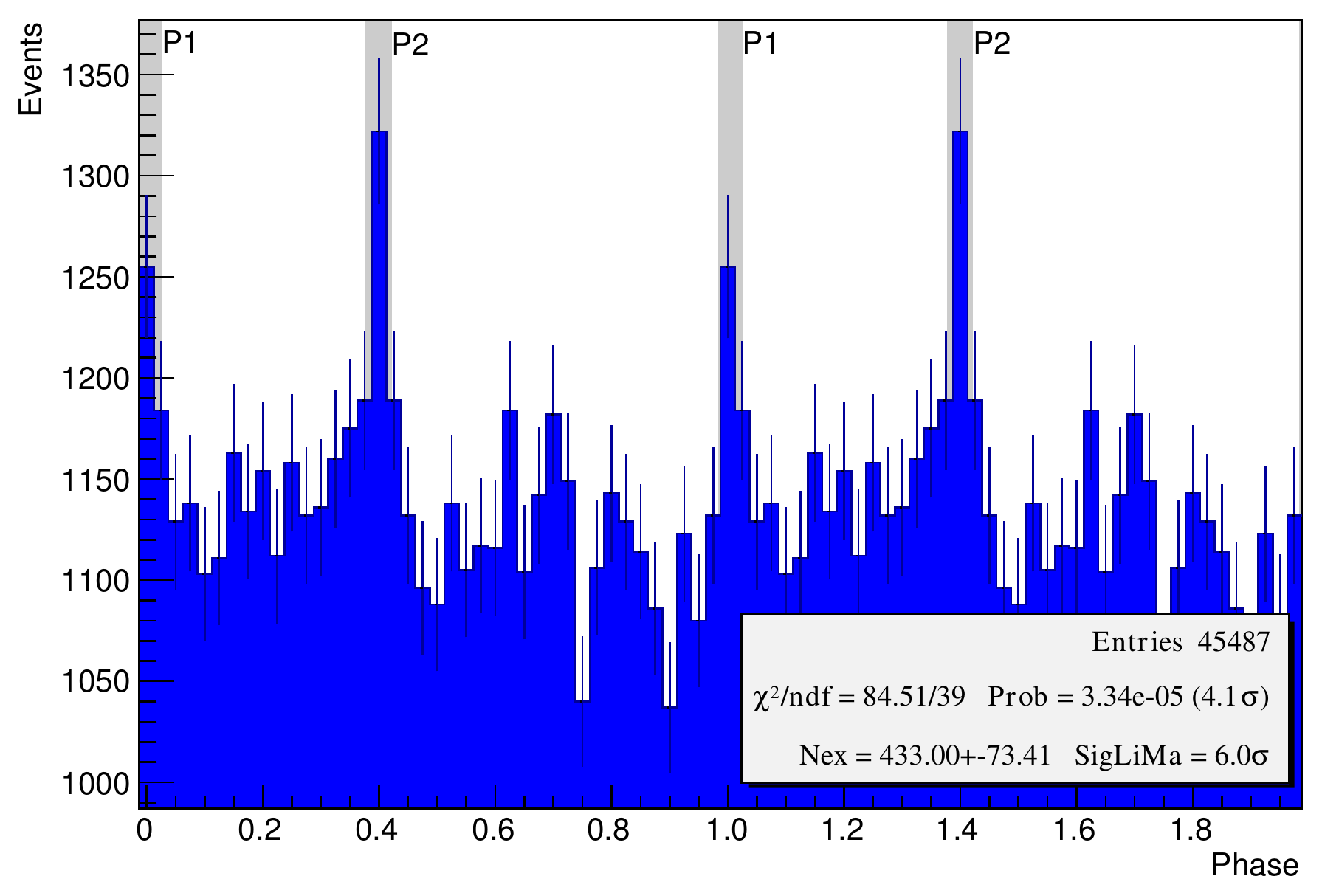}
   \caption{Phase diagram resulting from the MAGIC data after the barycentering process and the cut selection, in the energy range 43-368~GeV. The resulting significance is 6$\sigma$. The gray areas correspond with the pulsed emission regions as determined by \citet{2012A&A...540A..69A} for the energy range 50-400~GeV. P1 is visible around phase values of 0 and 1, whereas P2 is located around 0.4 and 1.4.}
              \label{fig:phasediagram_VHE}
    \end{figure*}

The radio observations were carried out with the Effelsberg radio telescope and the Westerbork Synthesis Radio Telescope (WSRT) at a frequency of about 1.4~GHz. The two facilities scheduled complementary observations in order to exclude overlaps in the recorded data sample. Observations of the Crab pulsar with the Effelsberg radio telescope were carried out in baseband mode with the P217~mm and P200~mm prime focus receivers and the PSRIX pulsar backend \citep{2016MNRAS.458..868L}. The Crab pulsar observations taken with the WSRT were carried out with 13 out of 14 available antennas, their Multi-frequency Front End Receivers (MFFEs, \citealt{1982ITMTT..30..201C,Tan1991}), and the PuMa II pulsar backend \citep{2008PASP..120..191K}. \par
All radio data sets were coherently dedispersed \citep{1975MComP..14...55H} during an off-line reduction process. For this part of the reduction the digital library DSPSR \citep{2011PASA...28....1V} was used. After the dedispersion procedure, the resulting data sets were phase folded with ephemeris files obtained from the Jodrell Bank Observatory \citep{1993MNRAS.265.1003L}. To ensure absolute alignment between the radio and $\gamma$-rays pulses, an ephemeris which covered the observing days was created and used instead of the monthly released one. To extract the brightest single pulses, an additional data selection, based on the standard deviation of the signal in the OFF-pulse radio emission regions, was introduced in the dedispersed data sets. With this technique, a total number of 99444 GPs was extracted from the radio data. \par
A summary of all the radio observations performed for this study is given in Table~\ref{table:summary_radio} and a corresponding phase diagram of an observation taken with the Effelsberg telescope and the WSRT is shown in Fig.~\ref{fig:phasediagram_radio}.

 \begin{figure}
   \centering
   \resizebox{\hsize}{!}{\includegraphics{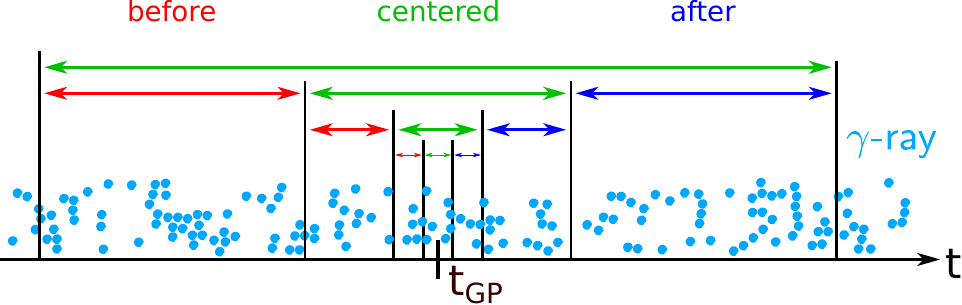}}
   \caption{Construction of search windows around a radio GP. The central window is symmetric around the arrival time of a radio GP. The advanced and delayed windows have the same length and are adjacent in time to the centered window. This construction arranges the search windows in a hierarchy where all three search windows of one length together form the centered search window of the next larger duration.
              }
              \label{fig:search_windows_radio}%
    \end{figure}

\begin{table*}
\caption{Summary of radio and VHE $\gamma$-ray observations.}
\label{table:summary_radio}
\centering
\begin{tabular}{r r r r r r r r}
\hline\hline
Epoch & $\nu$  & BW  & Facility & N$_{GP}$ & T$_{radio}$  & T$_{MAGIC}$  & T$_{overlap}$ \\ {[YYYY-MM-DD]} & [MHz] & [MHz] & & & [h] & [h] & [h]\\ 
\hline
  
  2012-12-07 & 1347.5   & 200 & Eff     & 1687      &  1.8 & 2.9  & 0.8 \\
  2012-12-10 &1380.0 &  160  & WSRT & 15456   & 2.0  &  2.0   & 1.9 \\
  2012-12-17 &  1347.5 & 200  & Eff & 3429          & 2.3 &  1.7   & 1.6  \\
  2013-01-08  &  1380.0  & 160 & WSRT  & 5058  & 0.4 &  1.8   & 0.3  \\
  2013-01-09  & 1372.5 & 200  & Eff  & 3525         & 1.7 &  1.9   & 1.2  \\
  2013-01-10  &  1380.0 & 160 & WSRT  & 24274 & 2.0 &  1.8   & 1.5  \\
  2013-01-12  &  1347.5 & 200 & Eff  & 6445          & 2.8 &  1.8   & 1.4 \\
  2013-01-31  &  1347.5  & 200 & Eff  & 1688        & 1.5 &   2.0   &. 0.7  \\
  2013-02-02  & 1380.0  & 160 & WSRT  & 7118   & 0.9 &  2.0   & 0.9  \\
  2013-02-03  & 1380.0 & 160 & WSRT  & 18392  & 2.0 &  1.8   & 1.7  \\
  2013-02-06  &  1347.5 & 200 & Eff  & 4470         & 2.2 &  1.3   & 0.1  \\
  2013-02-07  &  1410.0 & 75 & Eff  & 696             & 1.2 &  1.7   & 1.1  \\
  2013-02-08  &  1347.5  & 200 & Eff  & 1046        & 0.5 &  1.7   & 0.1  \\
  2013-02-09  & 1347.5  & 200 & Eff  & 3821        & 2.0 &  2.0   & 1.4  \\
  2013-02-10  & 1347.5 & 200 & Eff  & 2339          & 1.7 &  1.8  & 1.3 \\
  \hline
  Total & & & & 99444 & 24.8 & 28.3 & 16.0  \\
\hline
\end{tabular}
\tablefoot{The value $\nu$ stands for the center frequency, BW for the bandwidth, N$_{GP}$ for the number of extracted GPs, T$_{radio}$ and T$_{MAGIC}$ indicate the duration of the radio and the corresponding VHE $\gamma$-rays observation respectively. The acronym Eff stands for Effelsberg radio telescope while WSRT for Westerbork Synthesis Radio Telescope.}
\end{table*}

\subsection{$\gamma$-Rays Observations}
VHE $\gamma$-rays observations were carried out with the MAGIC telescopes between December 2012 and February 2013 (simultaneously with observations either with the Effelsberg radio telescope, or the WSRT). They were taken at zenith angles of less than 30$\degr$ to achieve the lowest possible energy threshold, and with both telescopes in Wobble observation mode \citep{1994APh.....2..137F}. The reduction of the resulting data was carried out according to the standard analysis pipeline using the MAGIC Analysis and Reconstruction Software \citep[MARS,][]{zanin2013}. \par
To efficiently suppress the hadronic background without losing a large fraction of air showers induced by VHE photons from the Crab pulsar, energy dependent cuts in Hadroness (a test statistic for discrimination between a $\gamma$-ray or a hadron induced shower) and $\theta^{2}$ (the squared angular distance between the expected source position and the reconstructed one) parameters were performed \citep[details in][]{2016APh....72...76A}. They were optimized on an independent data sample of 46 hours of observations, taken at zenith angles of less than 30$\degr$, same as the main data set used in the present work. 
For an energy range spanning from 5~GeV to 50~TeV, 30 logarithmic energy bins were defined. In each energy bin the Hadroness and $\theta^{2}$ parameters were optimized to maximize the significance of the pulsed $\gamma$-ray signal taking into account the continuous emission from the Crab Nebula as described in \citet{2012A&A...540A..69A}.\par
After optimizing the cuts in each energy bin separately we picked the bins in the energy range from 42.9 to 367.8 GeV that correspond to the energy range in \citet{2012A&A...540A..69A}.
The 16 hours of VHE $\gamma$-ray data taken simultaneously with radio observations detected the pulsar clearly above the background of the Crab nebula (with 6.0 $\sigma$
significance) in that range, as shown in Fig.~\ref{fig:phasediagram_VHE}.

\begin{table*}
\caption{Comparison of the current data set with  previous MAGIC observations of the pulsed emission from the Crab Nebula.}
\label{table:compare_VHE_excess}
    \centering
    \begin{tabular}{c c c c c c c}
     \hline
     Reference & Emission Component & $N_{excess}$ & FWHM & $\sigma$ & Duration  & $E_{range}$\\
     & & & & & [h] & [GeV]\\
     \hline
     \hline
     Aleksi{\'c} et al. (2012) & P1+P2 & 1175 $\pm$ 116 & --- & 10.4 & 73 & 46-416 \\
     \hline
     &\multicolumn{6}{c}{}\\
     Aleksi{\'c} et al. (2014) & P1 & 930 $\pm$ 120 & 0.025 $\pm$ 0.007 & 8 & 135 & 50-400 \\
     \cline{2-7}
      & P2 & 1510 $\pm$ 120 & 0.026 $\pm$ 0.004 & 12 & 135 & 50-400 \\
     \hline
     &\multicolumn{6}{c}{}\\
     Ansoldi et al. (2016) & P1 & 1252 $\pm$ 442 & 0.010 $\pm$ 0.003 & 2.86 & 320 & 100-400 \\
     \cline{2-7}
      & P2 & 2537 $\pm$ 454 & 0.040 $\pm$ 0.009 & 5.66 & 320 & 100-400 \\
     \hline
    &\multicolumn{5}{c}{}\\
     & P1+P2 & 433 $\pm$ 73 & --- & 6.06 & 16 & 43-368 \\
    \cline{2-7}
     This work & P1 & 144 $\pm$ 41 & 0.015 $\pm$ 0.005 & 3.5 & 16 & 43-368 \\
     \cline{2-7}
     & P2 & 289 $\pm$ 58 & 0.036 $\pm$ 0.009 & 4.9 & 16 & 43-368 \\
    \hline
\hline
    \end{tabular}
\end{table*}
For the barycentering of the VHE $\gamma$-ray data the TEMPO2 pulsar timing software \citep{2006MNRAS.369..655H} and the same ephemeris files were used as for the radio data \citep{1993MNRAS.265.1003L}.
The folded light curve obtained after the barycentering process and the selection cuts is shown in Fig.~\ref{fig:phasediagram_VHE}: the gray shadowed areas are the results from the previous MAGIC phase resolved analysis of the Crab Pulsar \citep{2012A&A...540A..69A}. The overlap with the present data (blue filled area) shows the compatibility between the two analyses, even if the energy ranges for the two results are slightly different (our results are shown here for the energy range 43 to 368~GeV while results from \citet{2012A&A...540A..69A} were obtained in the energy range from 50 to 400~GeV. To further quantify the compatibility with previous MAGIC results, we report in Table~\ref{table:compare_VHE_excess} the number of excess events, significance and the full width at half maximum (FWHM) of a Gaussian fit to the peaks of P1 and P2 respectively, from the work of \citet{2012A&A...540A..69A,2014A&A...565L..12A}, \cite{2016A&A...585A.133A} and compare those with our results.

\section{Correlation Study}
\label{sec:correlation}
  \subsection{Approach}
Due to the lack of statistical methods for the correlation analysis of independent event lists\footnote{A discussion of that problem can be found in \cite{edelson_1988}.}, but also
for comparability with results from previous studies with IACT data, we adopted the approach described in \citet{2012ApJ...760..136A}. 
The number of coincidences between VHE $\gamma$-rays and GPs was %\nl{is}--I left was to be consistent with the time of the whole paragraph 
counted inside a given search window (SW, see Fig.~\ref{fig:search_windows_radio}). SWs were defined in terms of fractions or multiples of one rotational period of the Crab pulsar, namely: 1/9, 1/3, 1, 3, 9, 27, 81, 243, 729, 2187. 
With the aim of reducing the background emission from the Crab Nebula in our analysis, and to conduct dedicated studies on P1 and P2 respectively, we extended the approach of \citet{2012ApJ...760..136A} adopting SWs smaller than one rotational period (1/9 and 1/3).
Exploring different SWs allowed us to change the trade-off between statistical and systematic uncertainties. 
Moreover, the SWs smaller than one that we consider in the present work only contain one of the pulsed emission components, either P1 or P2, depending at which phase range the radio GPs are located.
Hence, SWs smaller than one rotation period of the Crab pulsar describe here the increase of VHE photons centered on radio GPs from only one of the regular emission components instead of both, and this allowed us to perform two separate analyses focused on P1 GPs and P2 GPs. As explained in Sec.~\ref{sec:intro}, the indication of different emission mechanisms of GPs in P1 and P2 makes the separated analysis an important tool to deeply investigate the possible coincidences between GPs at various phase ranges at different energies.\\ 
Since the emission mechanism of radio GPs is unknown, a delay in the generation of radio GPs and VHE photons cannot be excluded. Therefore the search windows were constructed for three different orientations in time: before, centered on, and after a radio GP (see Fig.~\ref{fig:search_windows_radio}). This way possible time delays between the generation of radio GPs and VHE $\gamma$ rays were included in the search procedure. The described approach results in a total of 30 correlation searches. 

\subsection{Monte Carlo Simulations}
\label{sec:MC_sim}
\subsubsection{Radio Simulations}
\label{sec:radio_MC}
Two statistical properties of the radio data were reproduced in the MC simulations: the average phase profile which we modeled by two Gaussians and the interarrival time between subsequent GPs. We modeled the interarrival times directly from the observed separations.\\
The interval between successive GPs was calculated and stored in a list. The list of interarrival times derived from observations was used instead of an analytic exponential distribution
for two reasons: 1) There were non-trivial deviations from the exponential distribution due to the phase bound occurrence of radio GPs; 2) There were deviations at large time separations (more than 50 rotation periods) due to the fact that the observations at both
telescopes were interrupted by weather, data write-out and other technical constraints. \\
Due to time gaps within the radio data sets (introduced during data recording to produce data chunks which were shorter in time and thus easier to reduce off-line), all interarrival times longer than 30 seconds were excluded from the simulation. All interarrivals shorter than this threshold were stored in a list. In the MC simulation a random interarrival time was fetched from the above-described list instead of drawing from an analytic exponential distribution. The parameters of the average profile were obtained by fitting  Gaussian distributions to the P1 and P2 components in the radio data. To increase the signal-to-noise-ratio, the fit was
performed on all the radio data collected during this campaign, with the exception of the Effelsberg data sets from 2013-01-09 and 2013-02-07 since both were taken at different center frequencies (see Table~\ref{table:summary_radio}). A more detailed explanation can be found in \citet{lewandowska_2015}.
\subsubsection{$\gamma$-Ray Simulations}
\label{sec:no-corr_sim}
In order to asses the significance level of the correlation, we produced correlation-free $\gamma$-ray data and searched for a correlation with the real radio data.
The \textit{synthetic} data had to reflect all the statistical properties of the real data. We produced such a data set in the following way\footnote{Additional details can be found in Appendix A.}:
\begin{enumerate}
     \item The rate of events (before selection by hadroness or $\theta^{2}$ parameters) was converted into a cumulative distribution function (CDF) with bin widths of one second. \item A uniform random number was drawn and the first bin in the CDF was located where the fraction of events exceeds that random number.
     \item A second uniform random number was drawn in order to determine a time stamp in the one second interval covered by the bin. 
     \item The time stamp obtained in step 3 does not yet reflect the fact that the VHE $\gamma$-ray data contains the pulsations from the Crab pulsar. Therefore, the time stamp was slightly modified in the following way: the event stayed within the same pulsar period, but the phase inside the rotation was drawn from a model containing a uniform background and two Gaussian peaks. This model was obtained by fitting the observed pulsed profile (after hadronness and $\theta^{2}$ cuts). The adjusted phase was then converted back into a time value using the Taylor expansion formula (Equation 8.4 in \citealt{handbook_2012}).
     \item The steps 2, 3 and 4 were repeated $M$ times. For each MC data set, $M$ was drawn randomly from a Poisson distribution with a mean of $N_{proc}$, where $N_{proc}$ is the total number of events after hadronnes and $\theta^{2}$ cuts. This way, one can get a {\it synthetic} uncorrelated VHE $\gamma$-ray data set with $M$ events. 
\end{enumerate}

To calculate confidence intervals with sufficiently low statistical error, 200 different {\it synthetic} VHE $\gamma$-ray data sets were produced by repeating the above procedure. As shown below in Section~\ref{res}, we did not find a statistically significant correlation. Therefore we calculated and report upper limits to the degree of correlation. For this purpose we defined a correlation parameter $\kappa$, which is the fraction of $\gamma$-ray events arriving simultaneously to an observed GP. Using this parameter we also generated {\it synthetic} correlated $\gamma$-ray data sets with different values of the parameter $\kappa$. At first we generated a uncorrelated $\gamma$-ray signal using the described procedure, but the arrival times of $\kappa \cdot N_{pulse}$ events were replaced by randomly picked arrival times of radio GPs, $N_{pulse}$ being the number of detected pulsed events in the real VHE $\gamma$-ray data.

\subsection{Results}\label{res}
\begin{figure*}
   \centering
   \includegraphics[width=0.75\textwidth, angle =0]{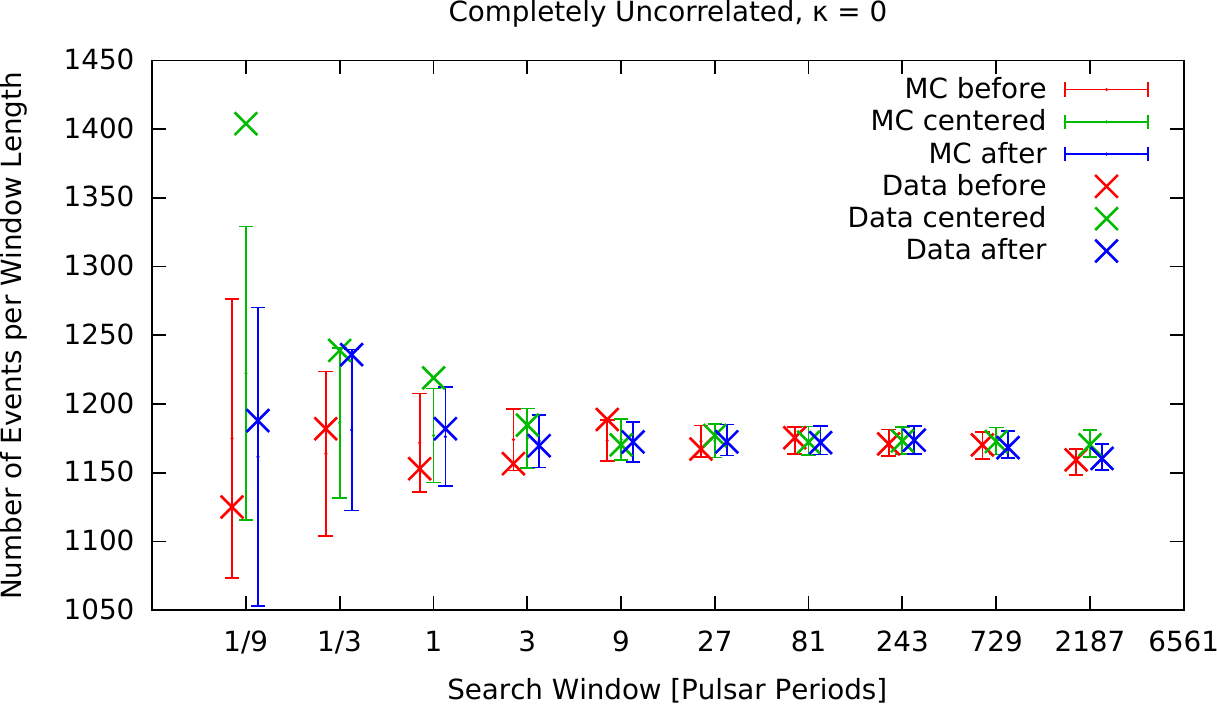}
   \includegraphics[width=0.75\textwidth, angle =0]{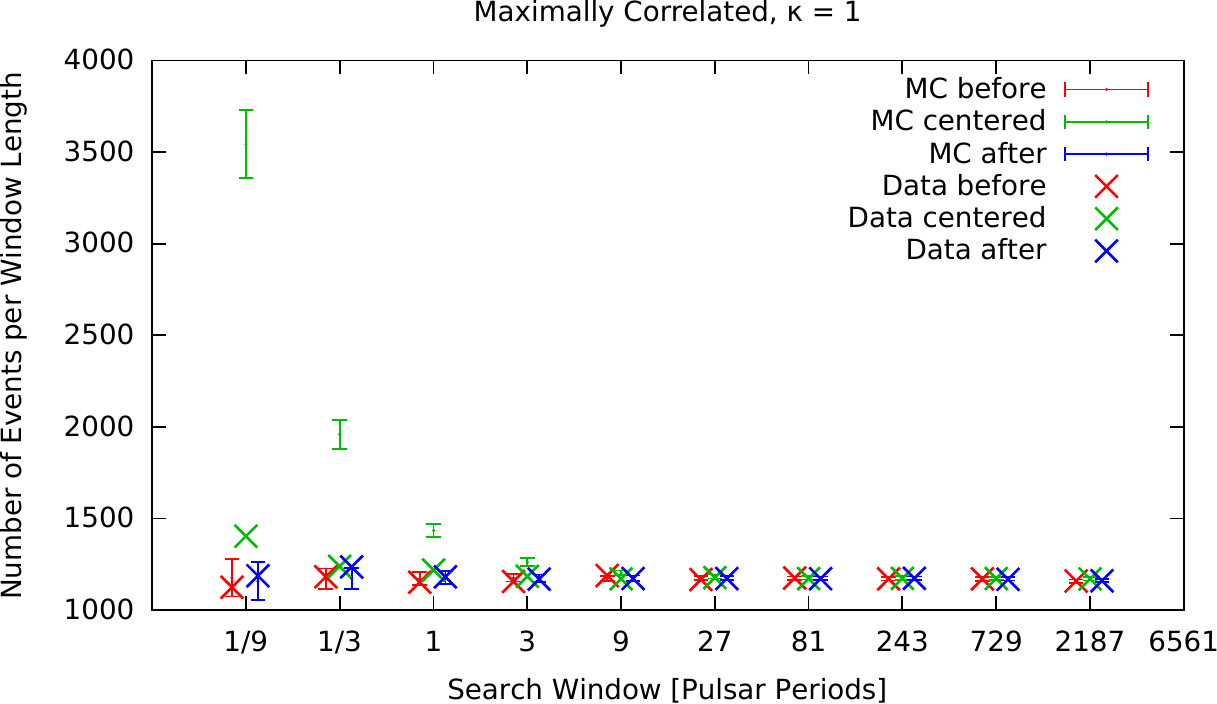}
   \caption{Enhancement of VHE photons around occurring radio GPs resulting out of data sets (marked with crosses) and VHE $\gamma$-ray MC simulations (indicated by bars). The upper plot shows the results for a perfectly uncorrelated VHE $\gamma$-ray signal in the MC simulations ($\kappa$ = 0), while the lower plot reports the flux enhancement results for an injected VHE $\gamma$-ray signal which is perfectly correlated with GPs resulting from the radio data in a centered search window ($\kappa$ = 1). The latter plot shows more clearly an increase of the number of VHE $\gamma$-rays centered on radio GPs for shorter search windows resulting from the data sets, indicating that the correlation is located at $\kappa$ < 1.}
              \label{fig:enhancement_MC_data}%
    \end{figure*}

 The number of coincidence events in the observational data for different SWs are shown in Fig.~\ref{fig:enhancement_MC_data}, together with the uncorrelated simulation results (top) and the perfectly correlated ($\kappa = 1$) simulation results (bottom).
 Error bars for simulation results are obtained as a 1 $\sigma$ fluctuation among 200 data sets (see Sec.~\ref{sec:MC_sim}).
 
\begin{figure*}
   \centering
   \includegraphics[width=8.5cm]{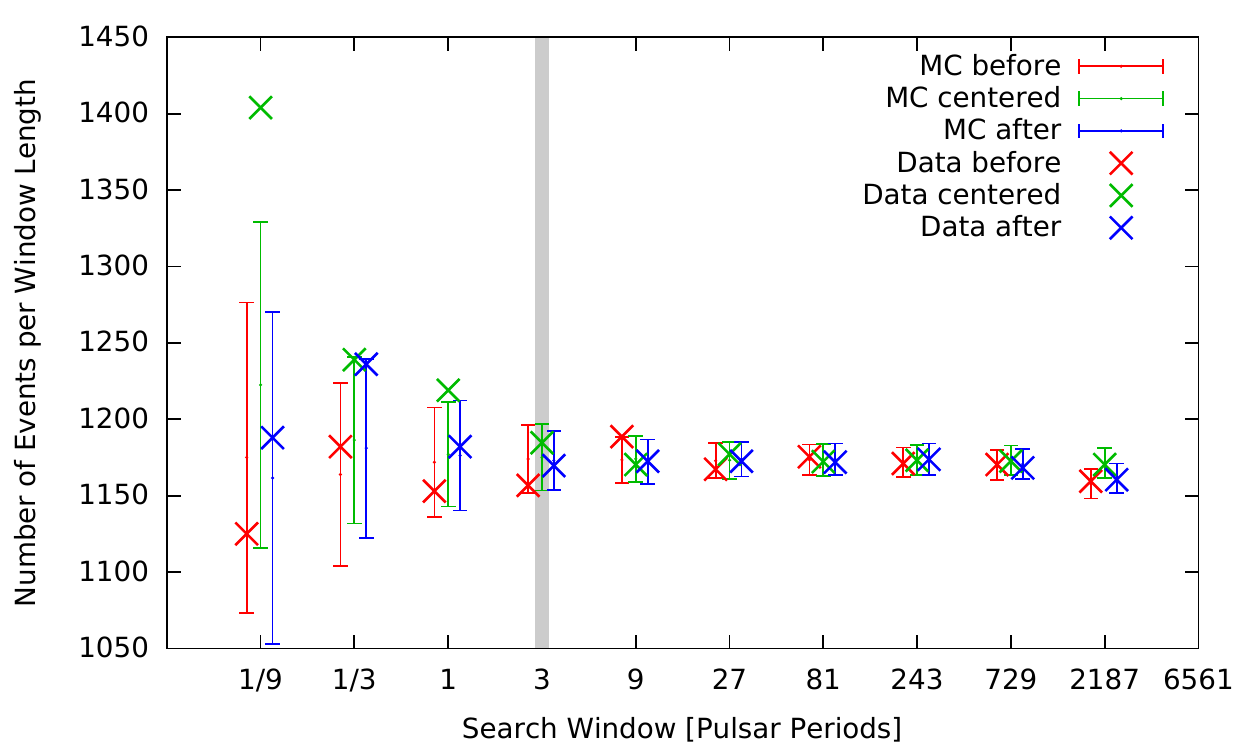}\includegraphics[width=8.5cm]{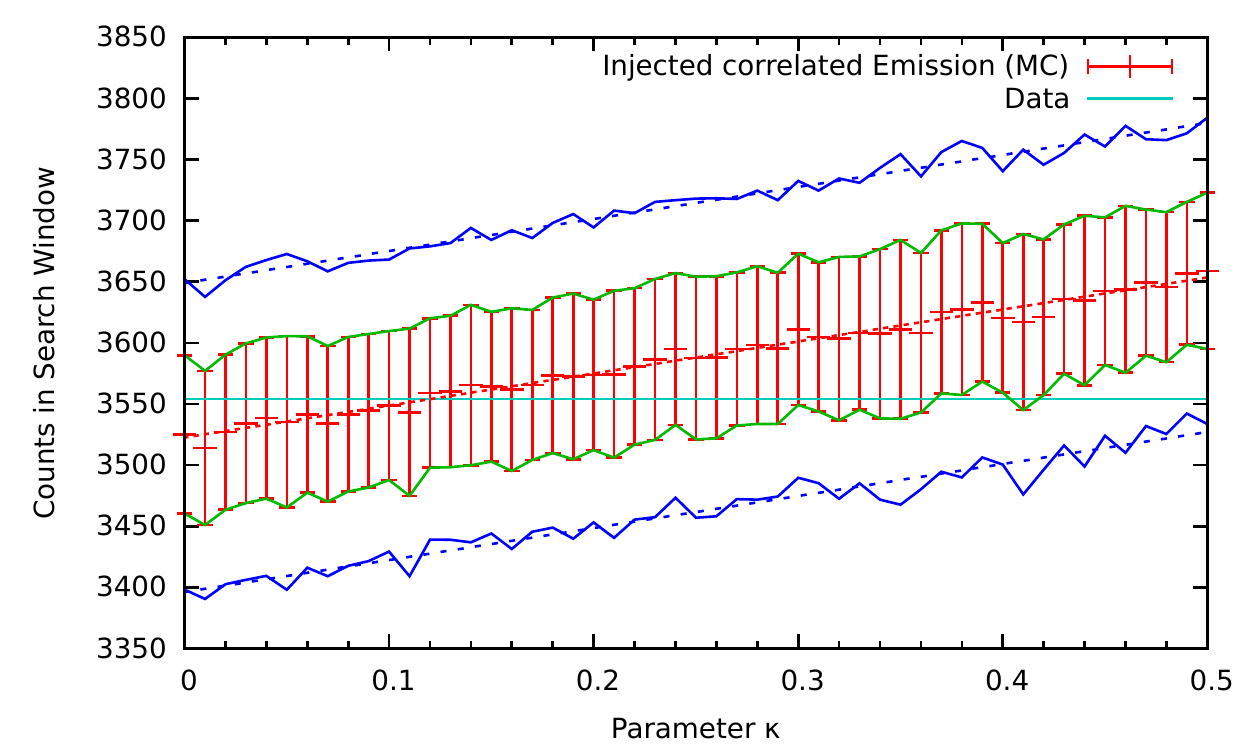}\\
  \includegraphics[width=8.5cm]{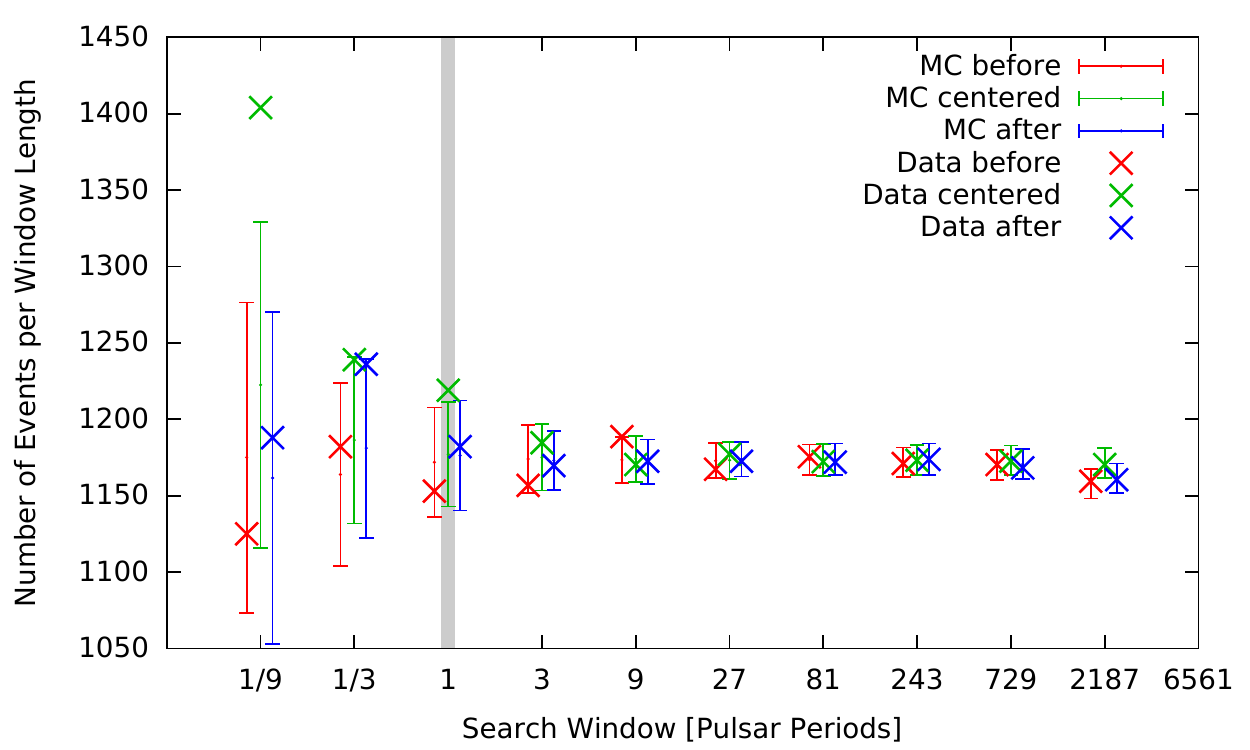}\includegraphics[width=8.5cm]{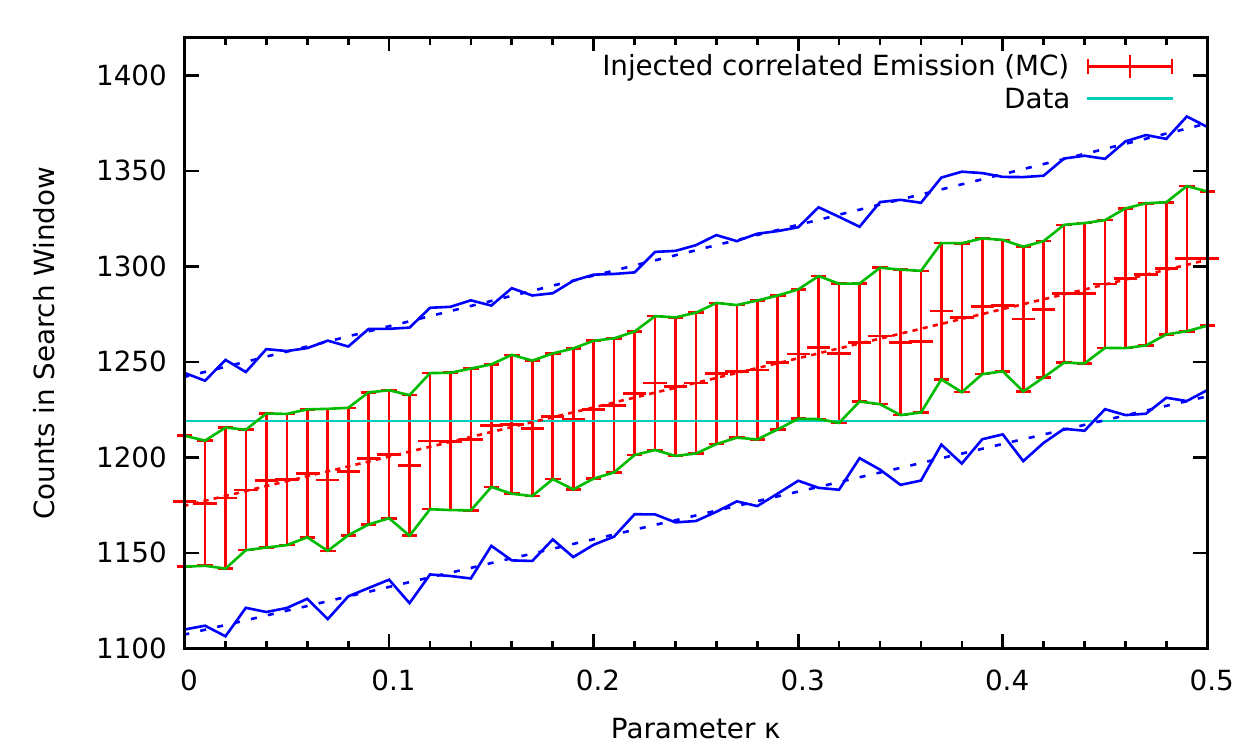}\\
  \includegraphics[width=8.5cm]{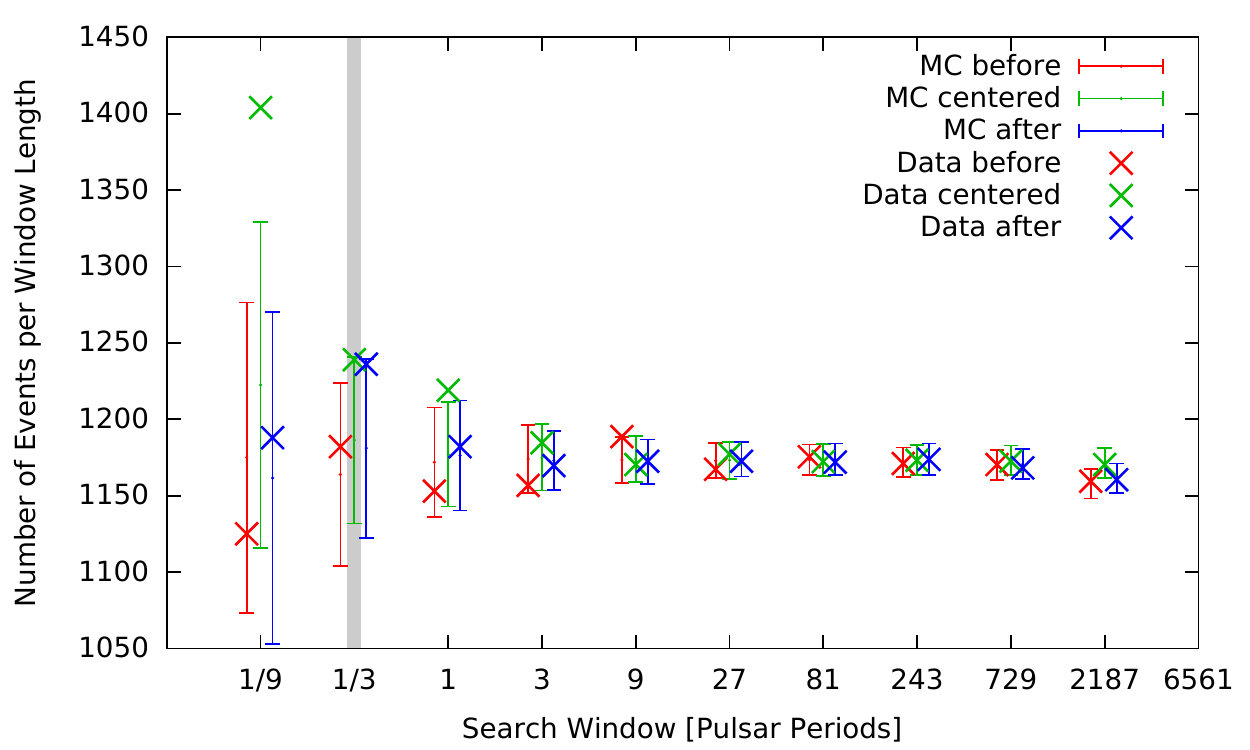}\includegraphics[width=8.5cm]{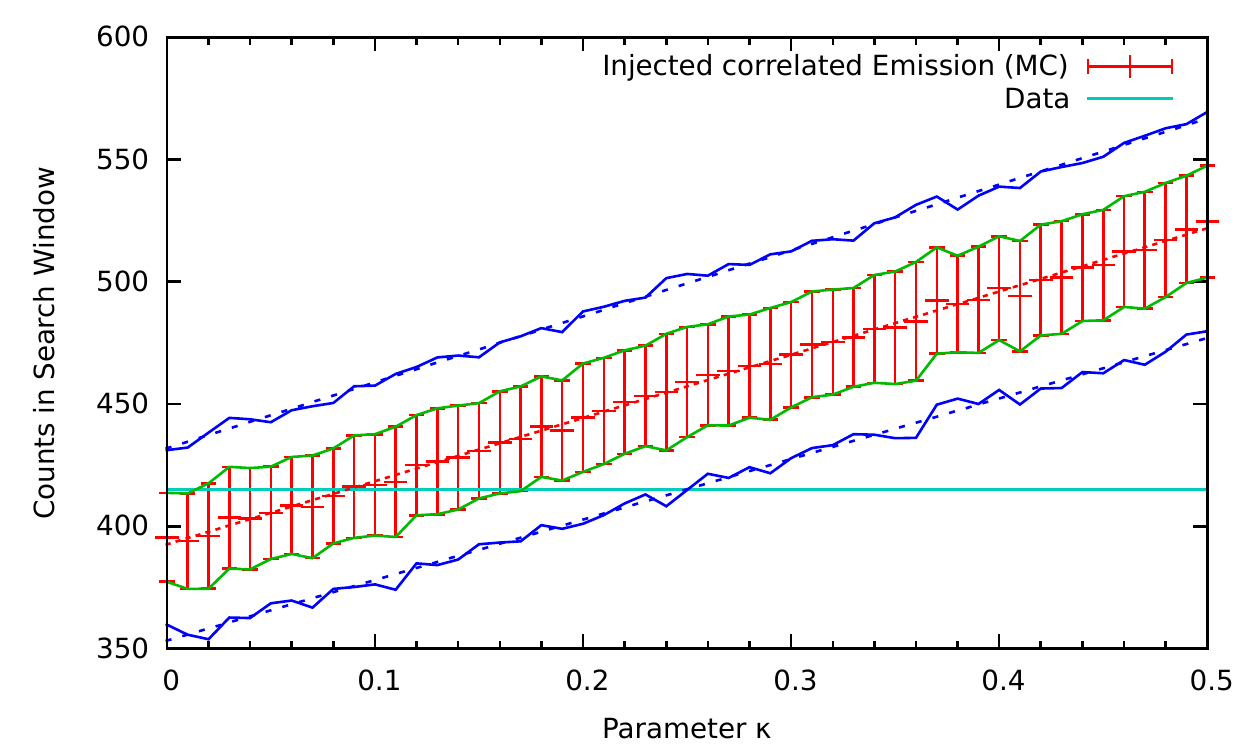}\\
  \includegraphics[width=8.5cm]{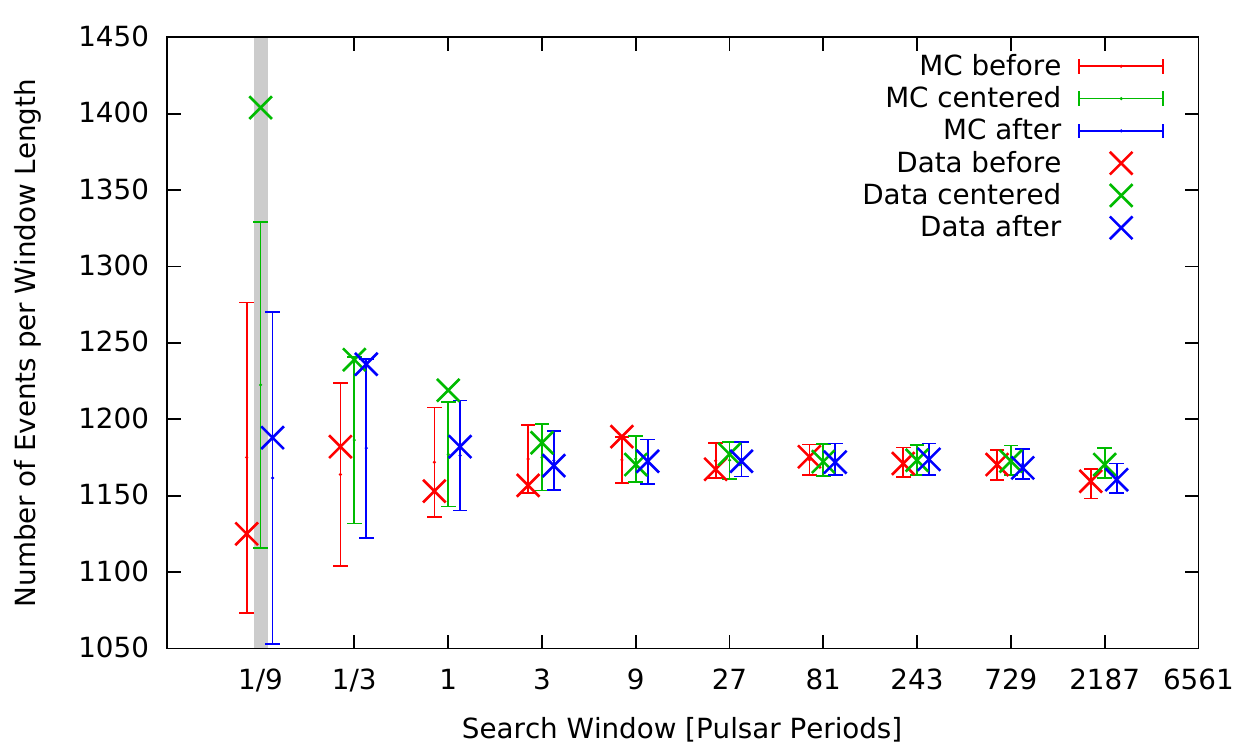}\includegraphics[width=8.5cm]{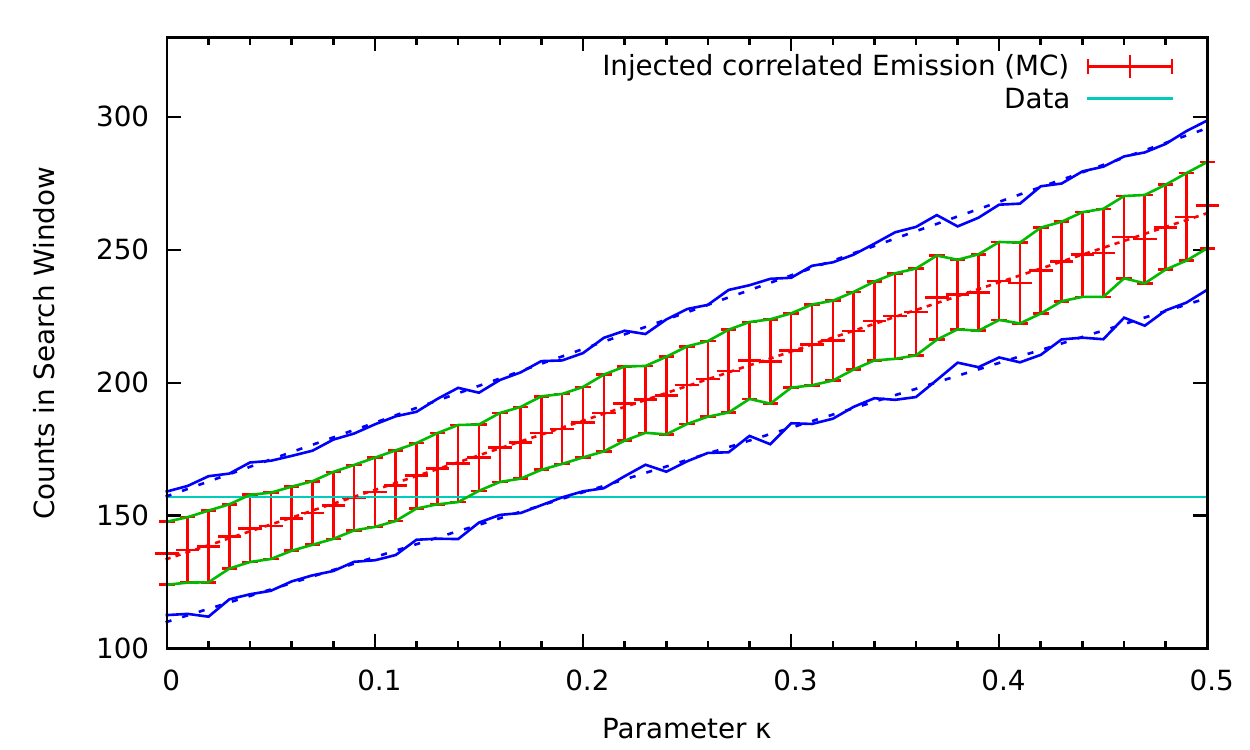}
   \caption{From top to bottom: Search windows of 3, 1, 1/3 and 1/9 Crab pulsar rotation periods length. Left: Enhancements of VHE $\gamma$-rays before, centered on and after a radio GP. The gray bar indicates the search window for which the $\kappa$ dependence is studied in the respective right hand plot. Right: The horizontal cyan line indicates the number of VHE $\gamma$-rays in the search window centered on GPs (the corresponding value from the observed data is indicated by a cross in the left hand plot. The normalization of the y-axis between the two columns is different). The plot also contains the results from 50 different sets of $\gamma$-ray MC simulations, using different values of $\kappa$. The average of each set is indicated by a red tick. The 1$\sigma$ range is indicated by the green lines, the 1.96$\sigma$ range (corresponding to a rejection of the null hypothesis on a $p = 0.05$ confidence level) is indicated by the blue lines. 
   }
              \label{fig:results_SW_ALL}%
    \end{figure*}

As can be deduced from the top panel of Figure \ref{fig:enhancement_MC_data}, the observed enhancement of VHE $\gamma$-rays becomes higher for shorter search windows centered on a GP, though the bottom panel shows the correlation is well below 100\%. Therefore, only the number of VHE photons in a search window centered on a radio GP for a window length of 1/9, 1/3, 1 and 3 Crab pulsar rotation periods will be examined in the forthcoming part of the analysis. To determine the enhancement quantitatively, MC simulations with different $\kappa$ values are compared with the corresponding data point. The results are shown in the right hand plots of Fig.~ \ref{fig:results_SW_ALL}. 

The cyan line represents the data point from the respective search window. The red ticks correspond to the average values of the VHE $\gamma$-ray MC simulations for different values of $\kappa$. The 1 $\sigma$ range around the average values is indicated by the vertical red bars as well as the green lines. The blue lines stand for the 1.96 $\sigma$ range. Since a linear scaling of both the average and the upper and lower limits are expected, the plot also contains fitted linear trend lines as dashed curves. To determine the best estimated value of $\kappa$ which reflects the enhancement of VHE $\gamma$-rays seen in the data, we calculate the intersection of the horizontal cyan data line with the linear trend lines\footnote{in one case (SW=3) this required extrapolation to $\kappa >$0.5, beyond the range for which MCMC simulations were performed.}. This procedure is carried out for all four search window lengths. The corresponding results are given in the upper 4 rows of Table \ref{table:kbest}.  The most significant deviation of $\kappa_{best}$ is seen at a  search window of 1/9 of the rotation period, which is in accordance with Figure \ref{fig:enhancement_MC_data}. 
Since none of the $\kappa_{best}$ is significantly larger than 0, 95\% level upper limits on $\kappa$ are also calculated on each window size, as the intersection between the cyan and blue lines in the figure. They are shown in the column of "CI$_{95\%}$" in Table \ref{table:kbest}.    
 
 \begin{table*}
\caption{Results of the correlation study between radio GPs and $\gamma$-rays that appear to be correlated with radio GPs (resulting from the intersection values between linear fits of $\gamma$-ray MC simulations and data points in the right hand figures of Fig.~\ref{fig:results_SW_MP} and Fig.~\ref{fig:results_SW_LFIP}).}
\label{table:kbest}
\centering
\begin{tabular}{r r r r r r}
\hline\hline
Component & SW  & $\kappa _{best}$ & CI$_{95\%}$ & p$_{err}$ & F$_{UL}$\\
& [P$_{crab}$] & & & & [\%]\\ 
\hline                        
  
ALL & 3& $0.12^{+0.23}_{-0.12}$ & 0.61 & 0.31  & 340\\ 
ALL & 1&  $0.17^{+0.14}_{-0.14}$ & 0.45 & 0.11 & 740\\
ALL & 1/3& $0.079^{+0.079}_{-0.079}$ & 0.24 & 0.14 & 1200\\
ALL & 1/9& $0.086^{+0.048}_{-0.048}$ & 0.19 & 0.04 & 2900\\
(P1+P2)$_{\gamma}$ & 3& $0.074^{+0.075}_{-0.074}$ & 0.24 & 0.19 & 12\\
(P1+P2)$_{\gamma}$ & 1& $0.018^{+0.043}_{-0.018}$ & 0.11 & 0.38 & 17\\
(P1+P2)$_{\gamma}$ & 1/3& $0.055^{+0.032}_{-0.055}$ & 0.13 & 0.06 & 60\\
P1$_{r}$& 3& $0.08^{+0.21}_{-0.08}$ & 0.52 & 0.37 & 15\\
P1$_{r}$& 1& $0.05^{+0.12}_{-0.05}$ & 0.30 & 0.35 & 25\\
P1$_{r}$& 1/3& $0.103^{+0.087}_{-0.087}$ & 0.30 & 0.10 & 76\\
P2$_{r}$ & 3& $0.129^{+0.055}_{-0.055 }$ &  0.25 & 0.01 & 34\\
P2$_{r}$ & 1& $0.018^{+0.029}_{-0.018 }$ & 0.09 & 0.27 & 36\\
P2$_{r}$ & 1/3& $0.016^{+0.022}_{-0.016}$ &  0.07 & 0.25 & 85\\
  
\hline
\end{tabular}
\tablefoot{The first column indicates the used data sample without phase cuts (marked "ALL"), only P1 and P2 in the VHE $\gamma$-ray data + MCs (marked with a $\gamma$) and P1, P2 based on Gaussian fits in the radio data (marked with an "r"). Their indices reflect whether the phase cuts are based on the radio, or $\gamma$-ray data. The acronym SW is standing for search window length, $P_{Crab}$ is the rotation period of the Crab pulsar, $\kappa_{best}$ is the intersection value, CI$_{95\%}$ is the upper value of the corresponding 95\% confidence interval, $p_{err}$ the probability to obtain the observed number of events based on the mean and standard deviation of the MC simulations and F$_{UL}$ the upper limit of the flux normalized to the pulsed VHE flux of the Crab pulsar.}
\end{table*}

\subsubsection{Phase-resolved Analysis}
\begin{figure*}
   \centering
   \includegraphics[width=8.5cm]{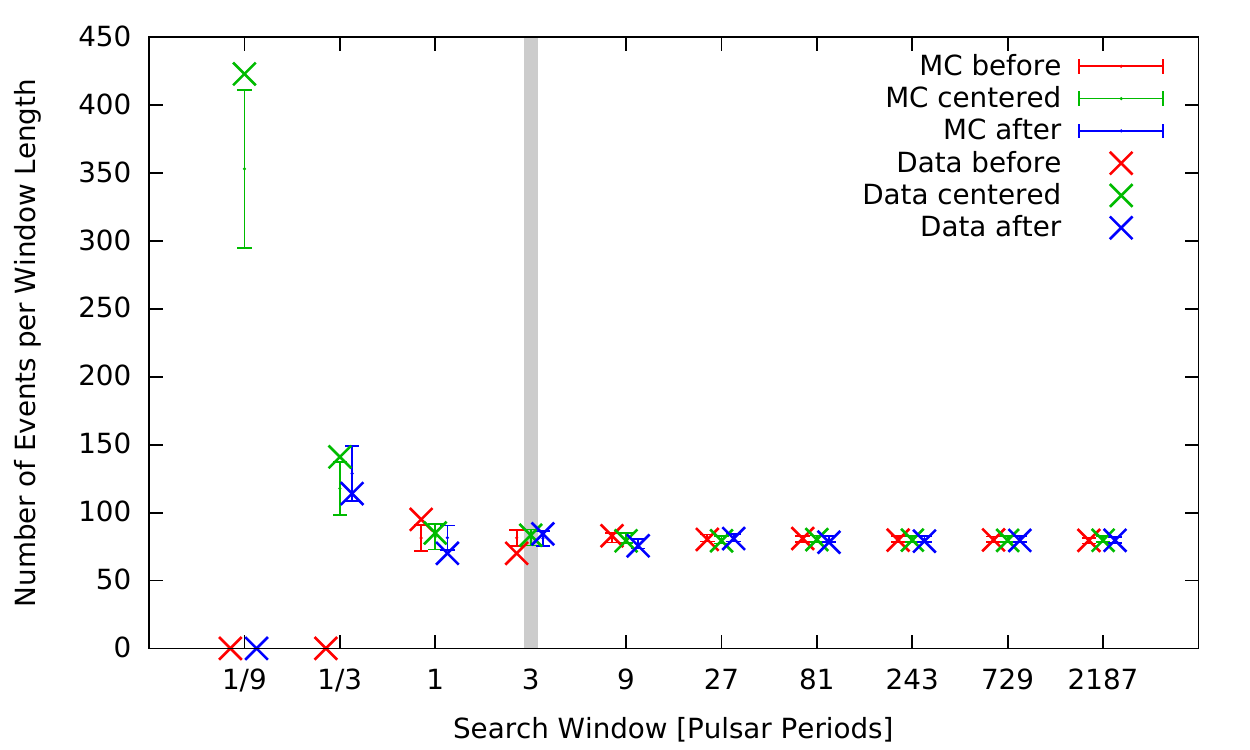}\includegraphics[width=8.5cm]{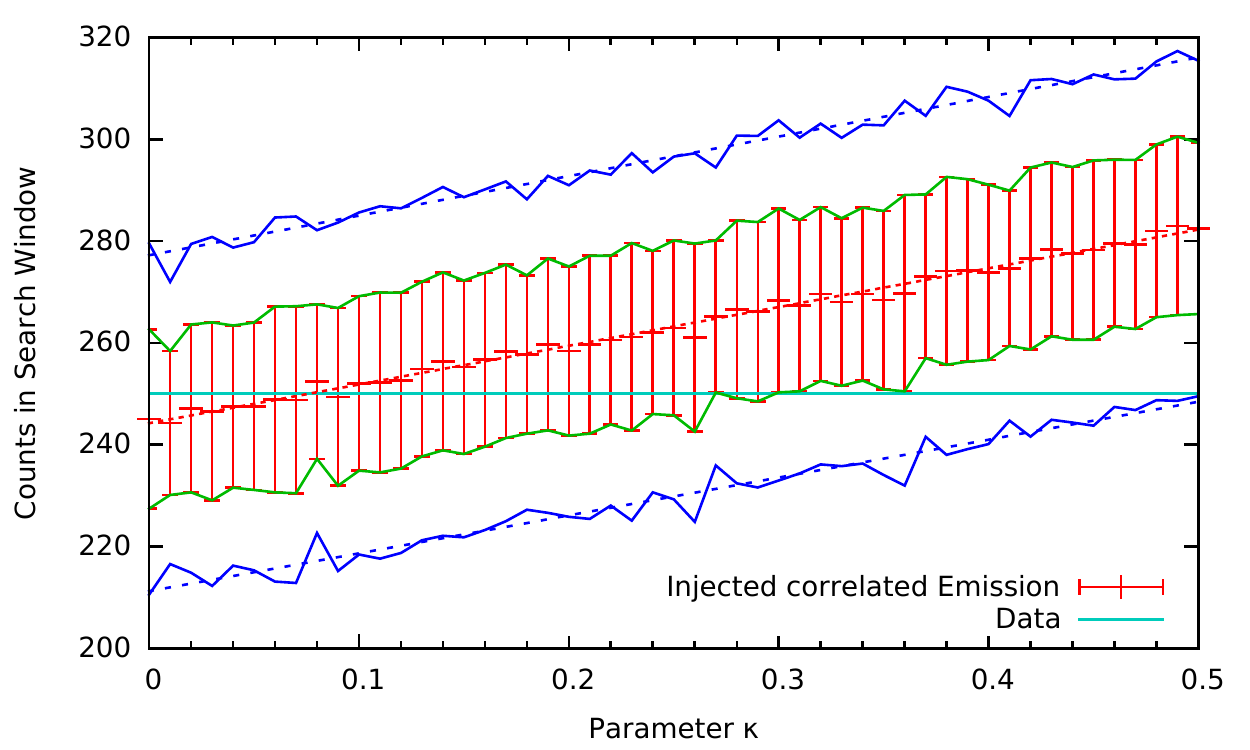}\\
  \includegraphics[width=8.5cm]{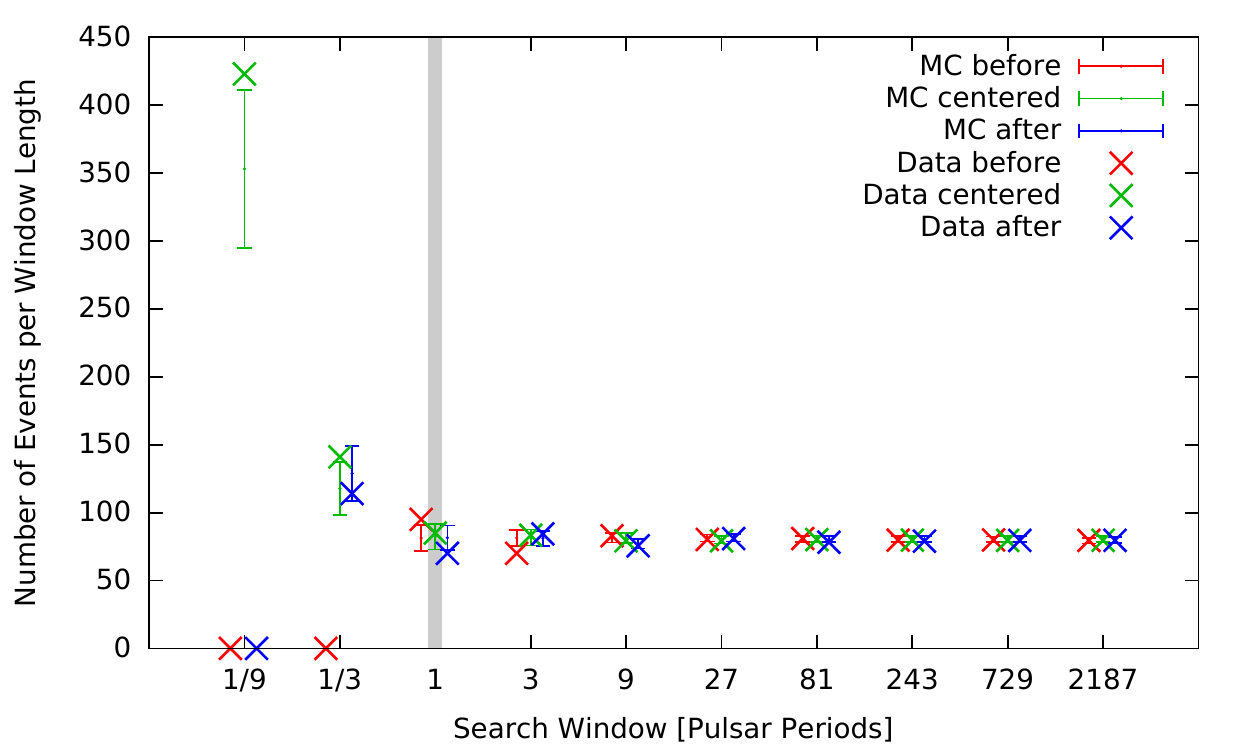}\includegraphics[width=8.5cm]{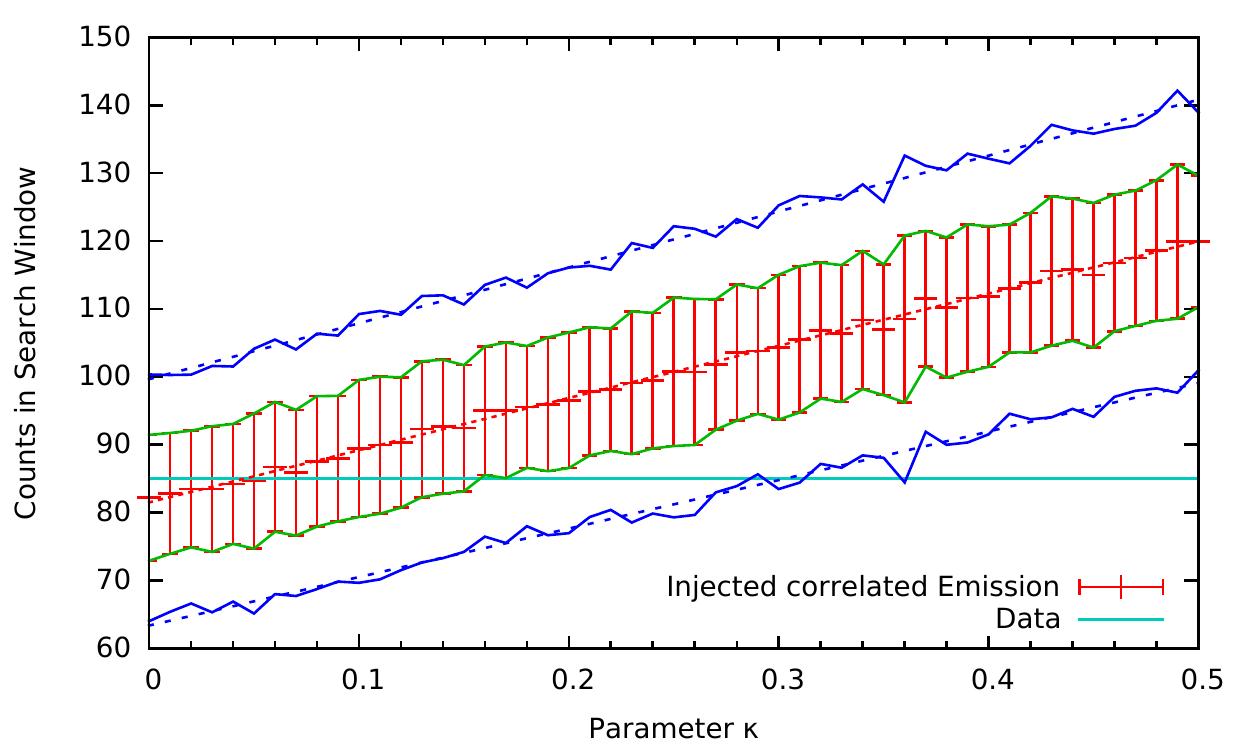}\\
  \includegraphics[width=8.5cm]{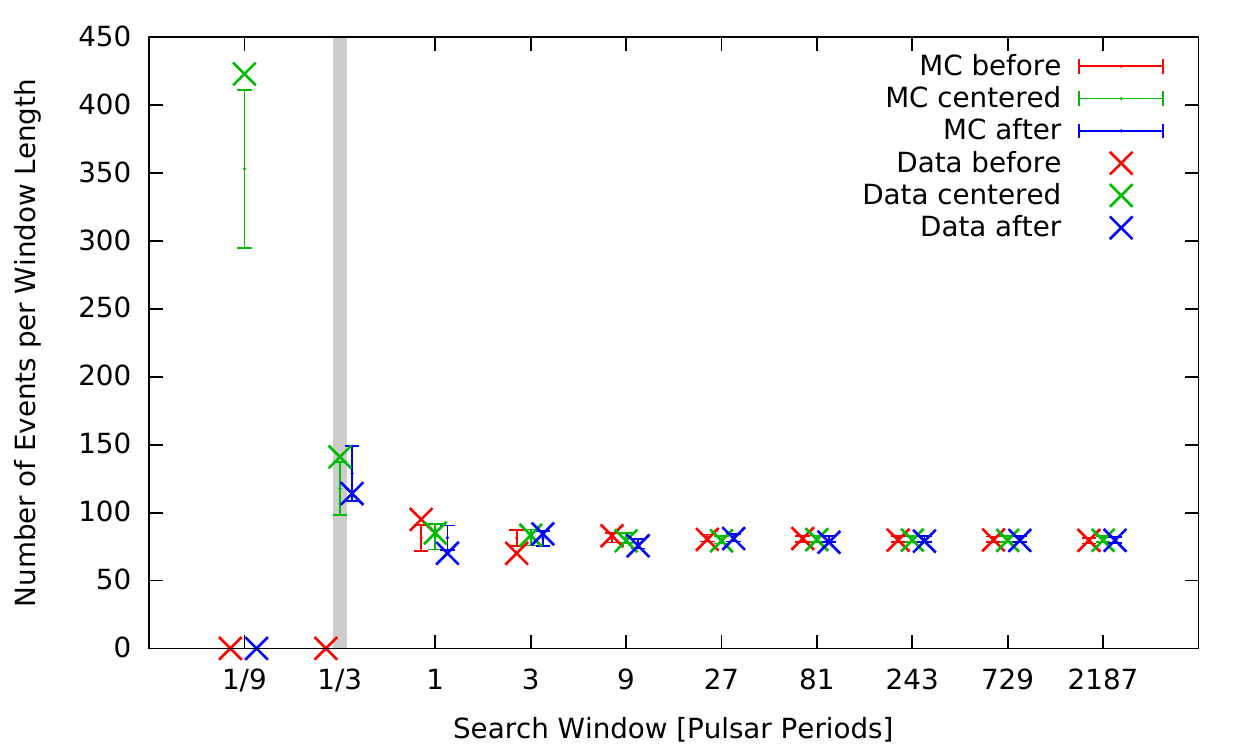}\includegraphics[width=8.5cm]{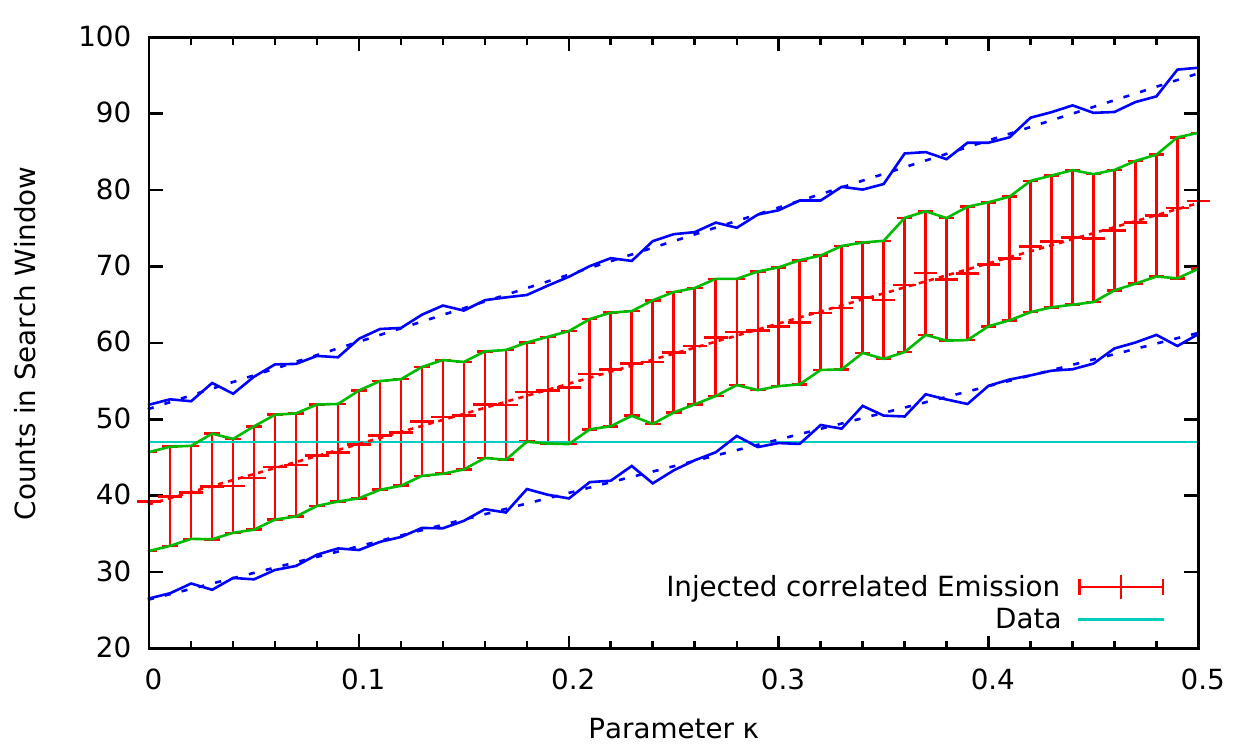}\\
  \includegraphics[width=8.5cm]{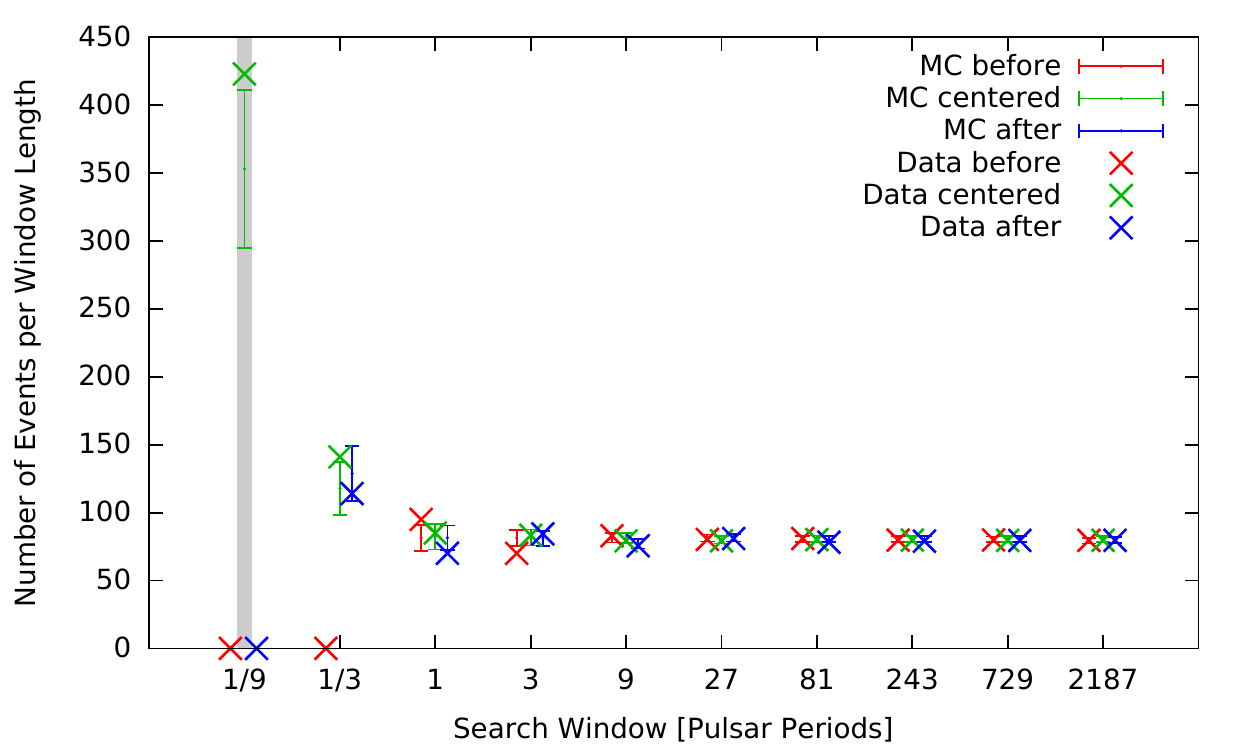}\includegraphics[width=8.5cm]{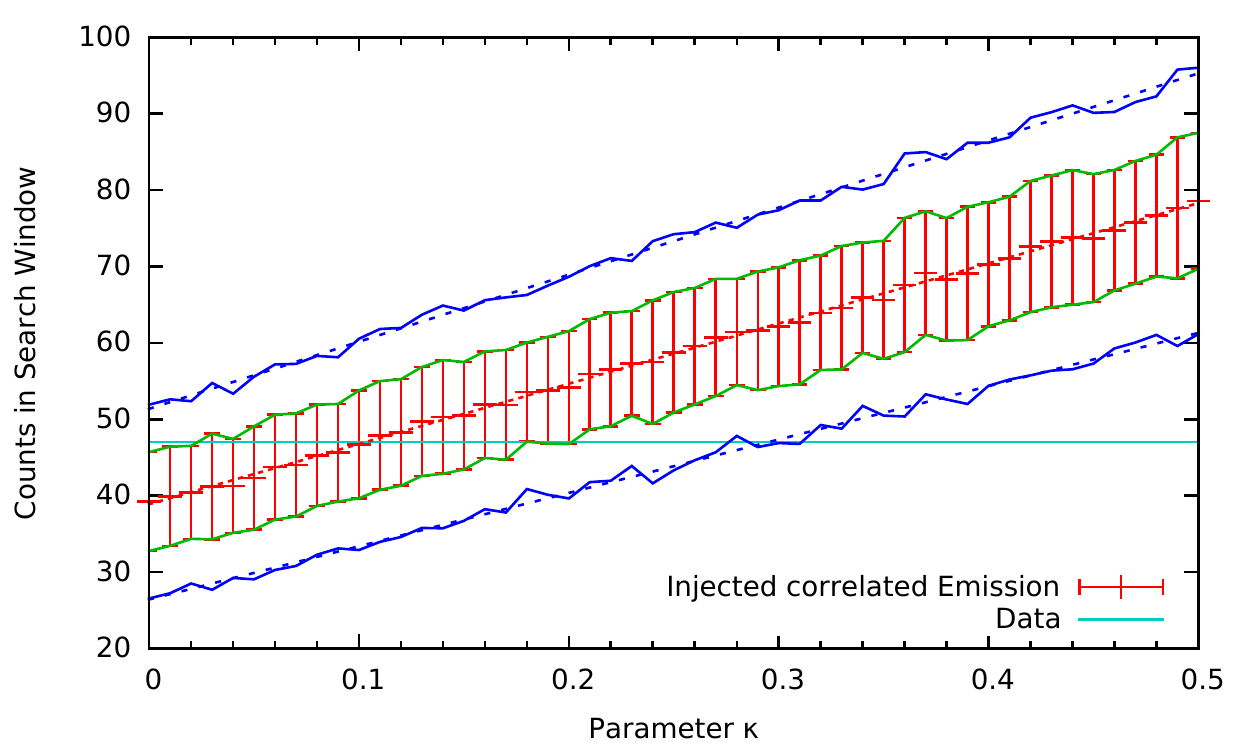}\\
   \caption{Results as described in Figure~\ref{fig:results_SW_ALL} for the P1 emission component (MP radio phase). The MC error bars in the left hand part of this figure were computed for $\kappa$ = 0 and not for the best fit $\kappa$ value.}
              \label{fig:results_SW_MP}%
    \end{figure*}

\begin{figure*}
   \centering
   \includegraphics[width=8.5cm]{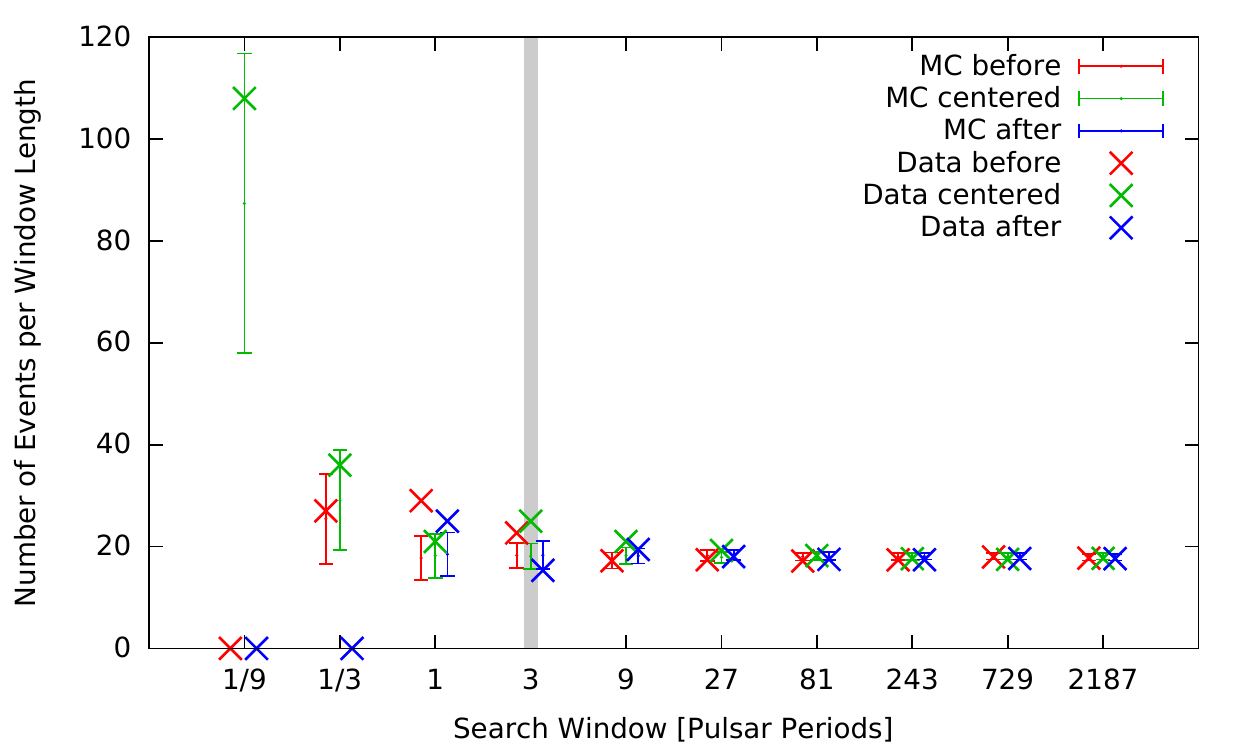}\includegraphics[width=8.5cm]{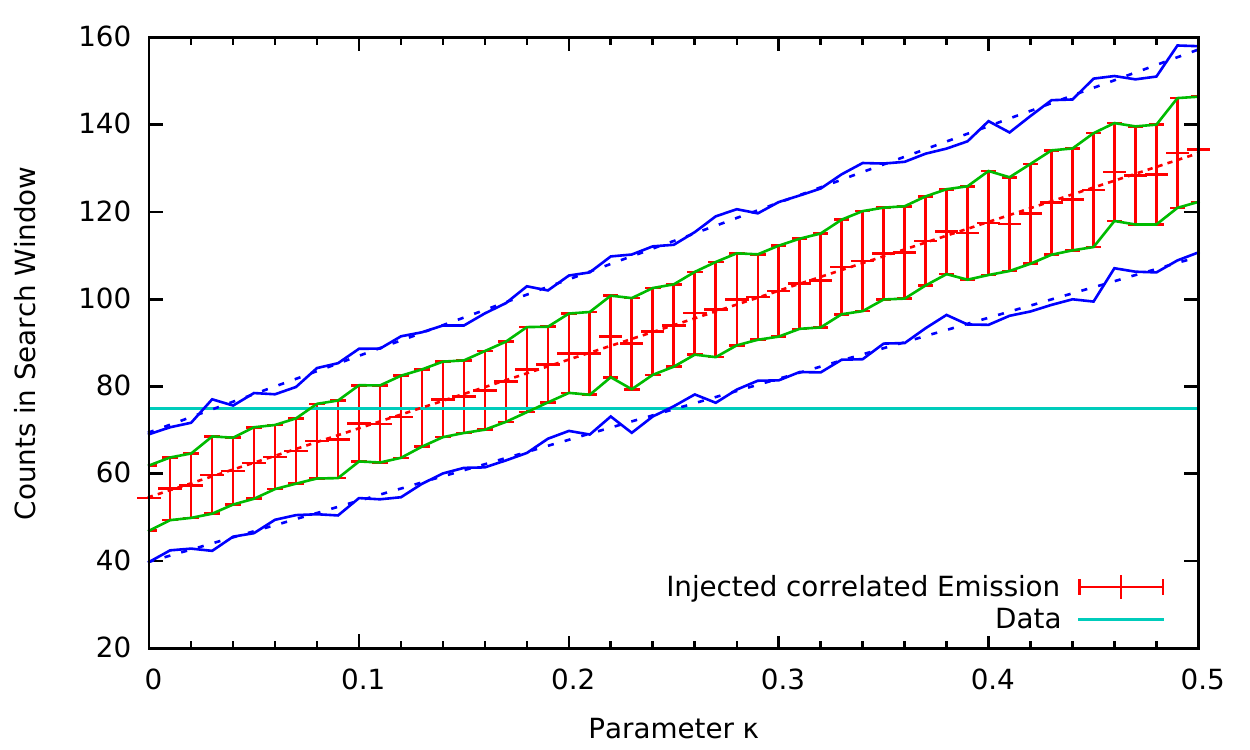}\\
  \includegraphics[width=8.5cm]{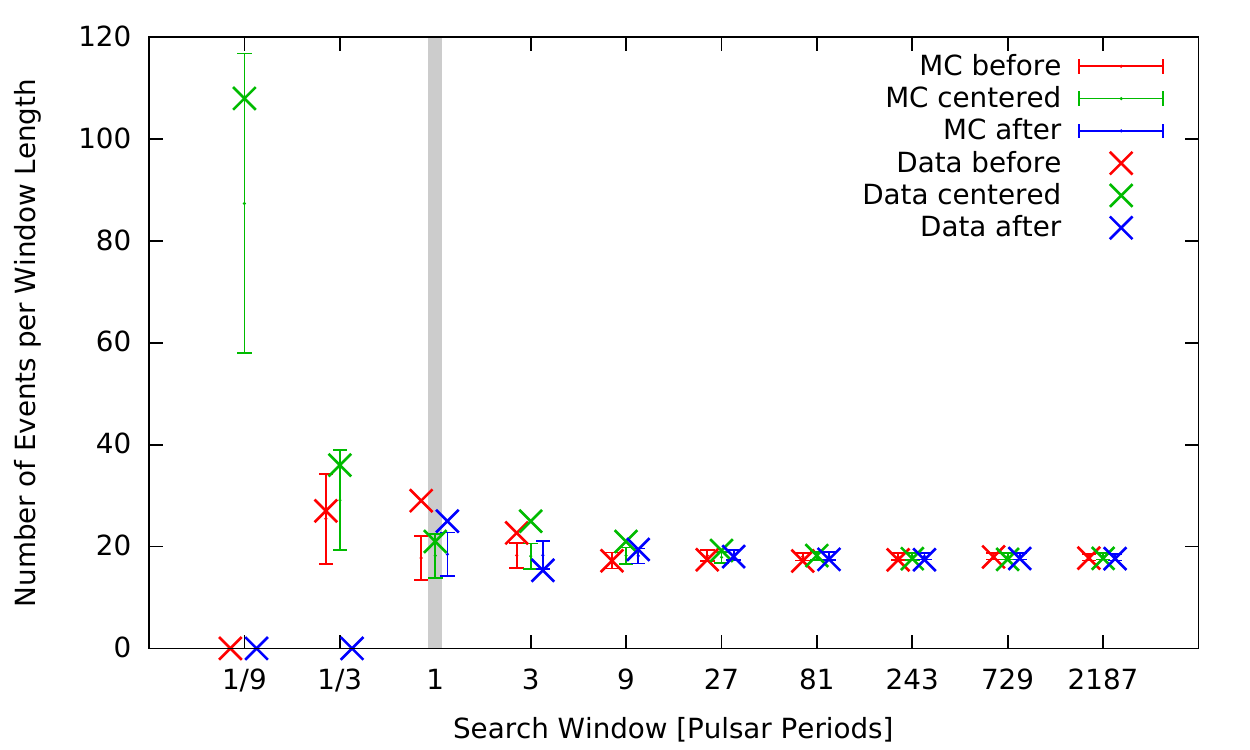}\includegraphics[width=8.5cm]{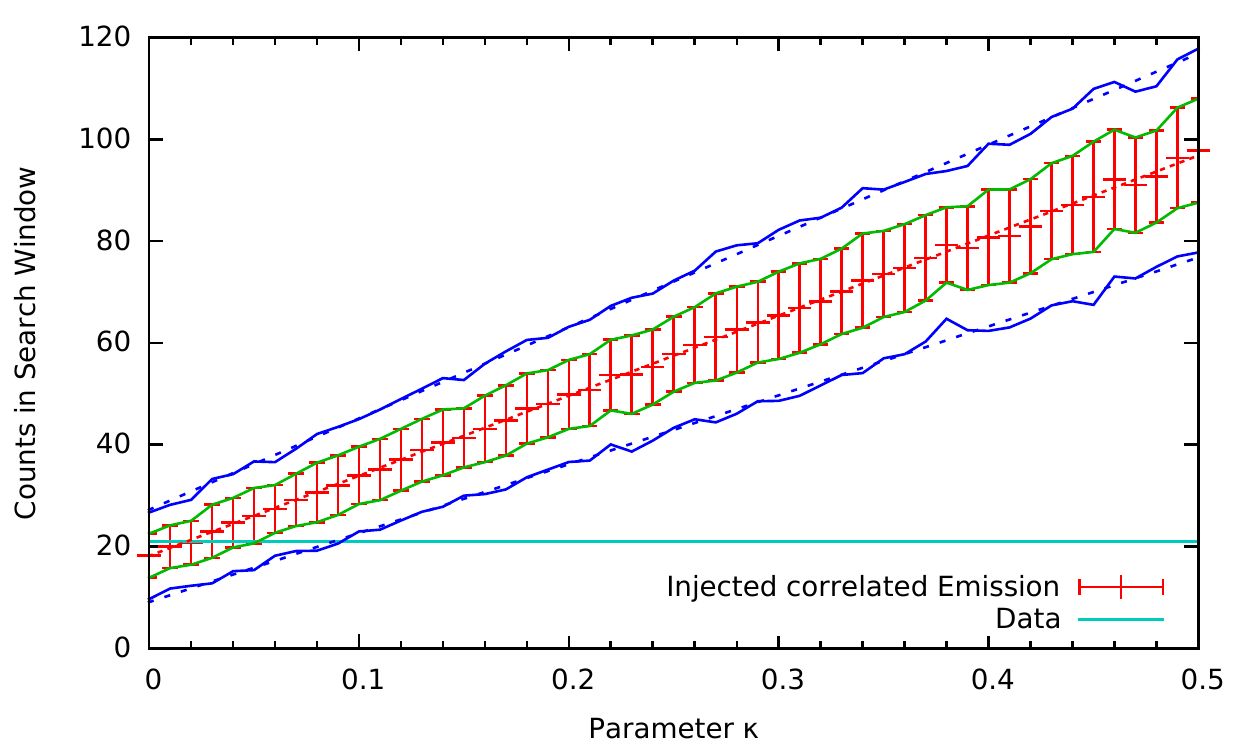}\\
  \includegraphics[width=8.5cm]{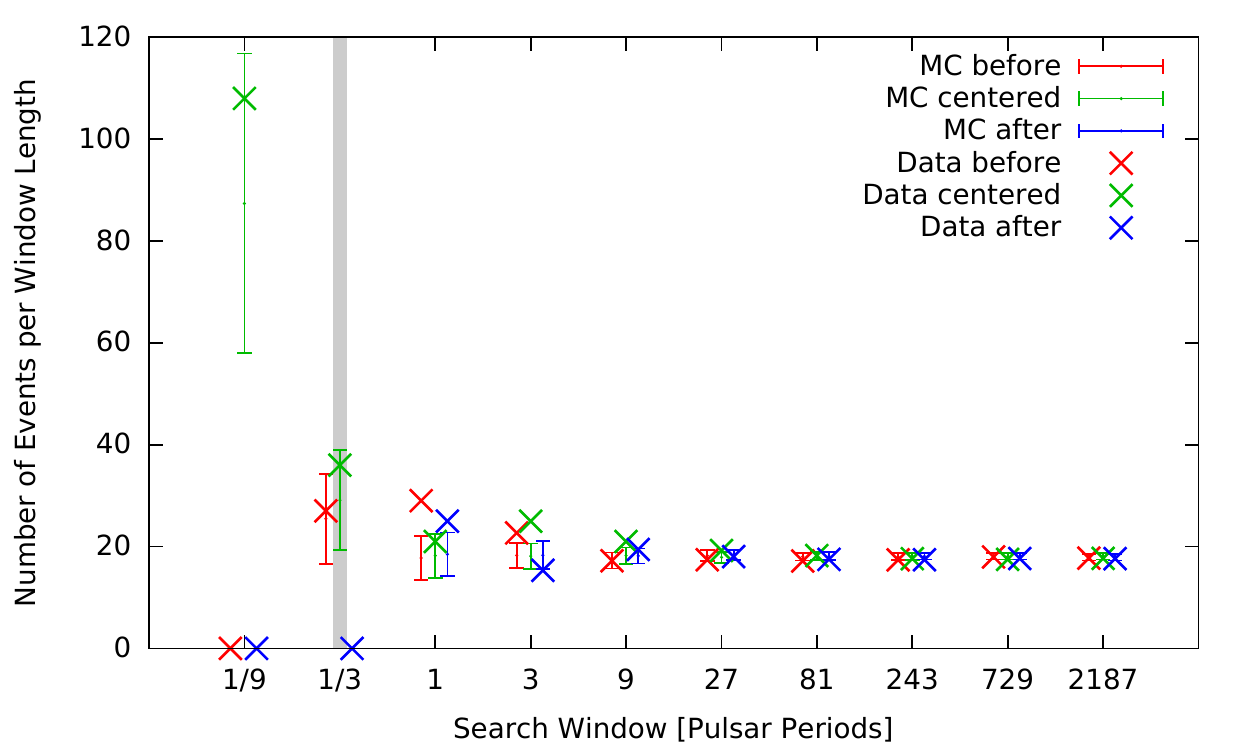}\includegraphics[width=8.5cm]{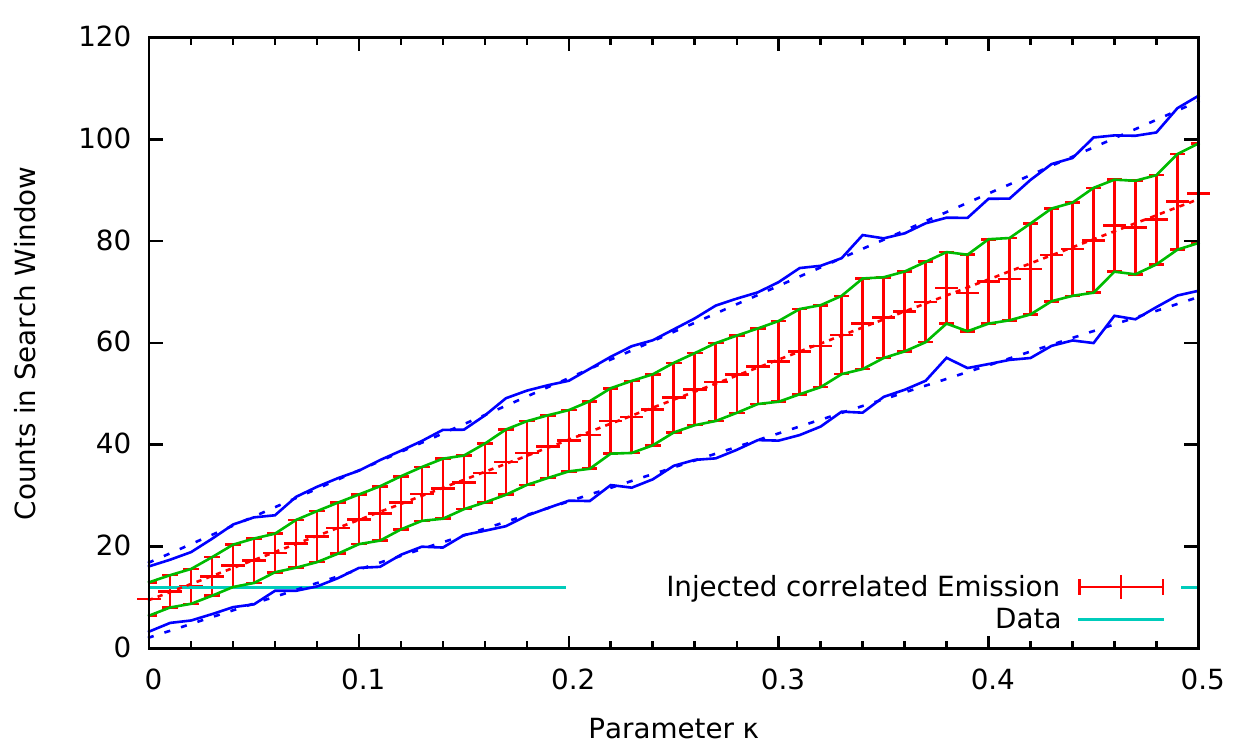}\\
  \includegraphics[width=8.5cm]{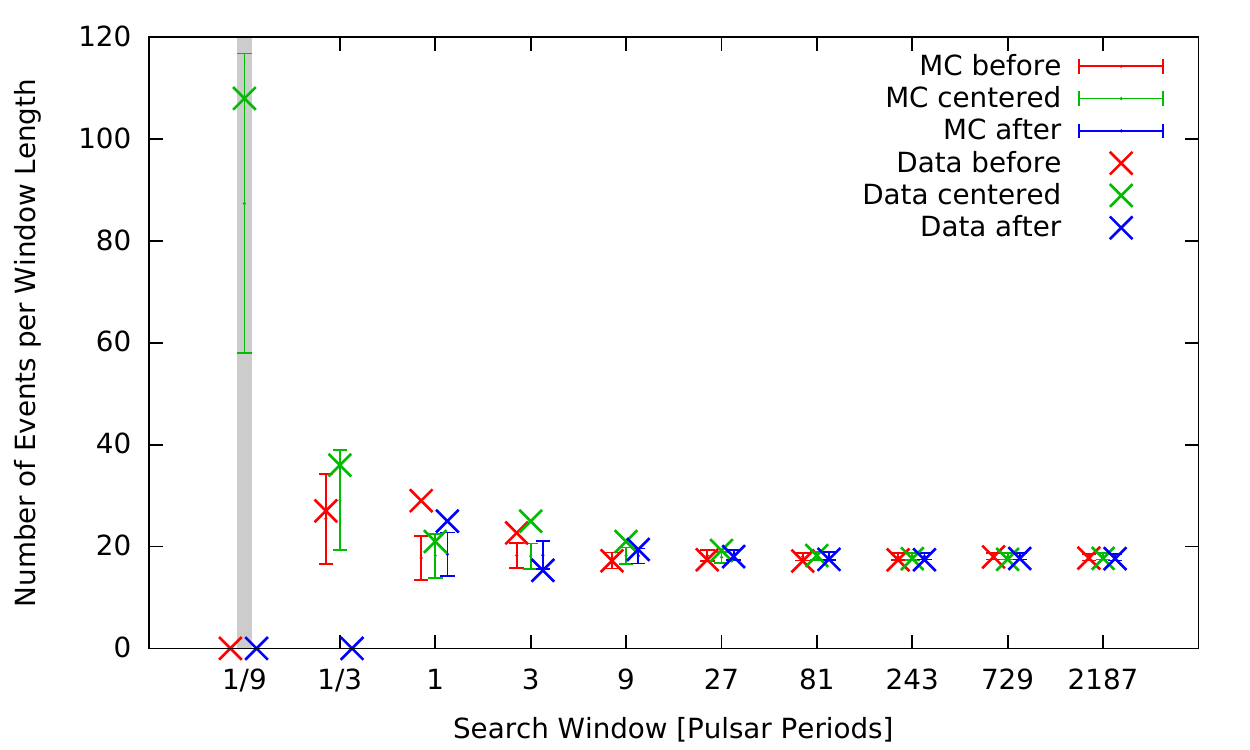}\includegraphics[width=8.5cm]{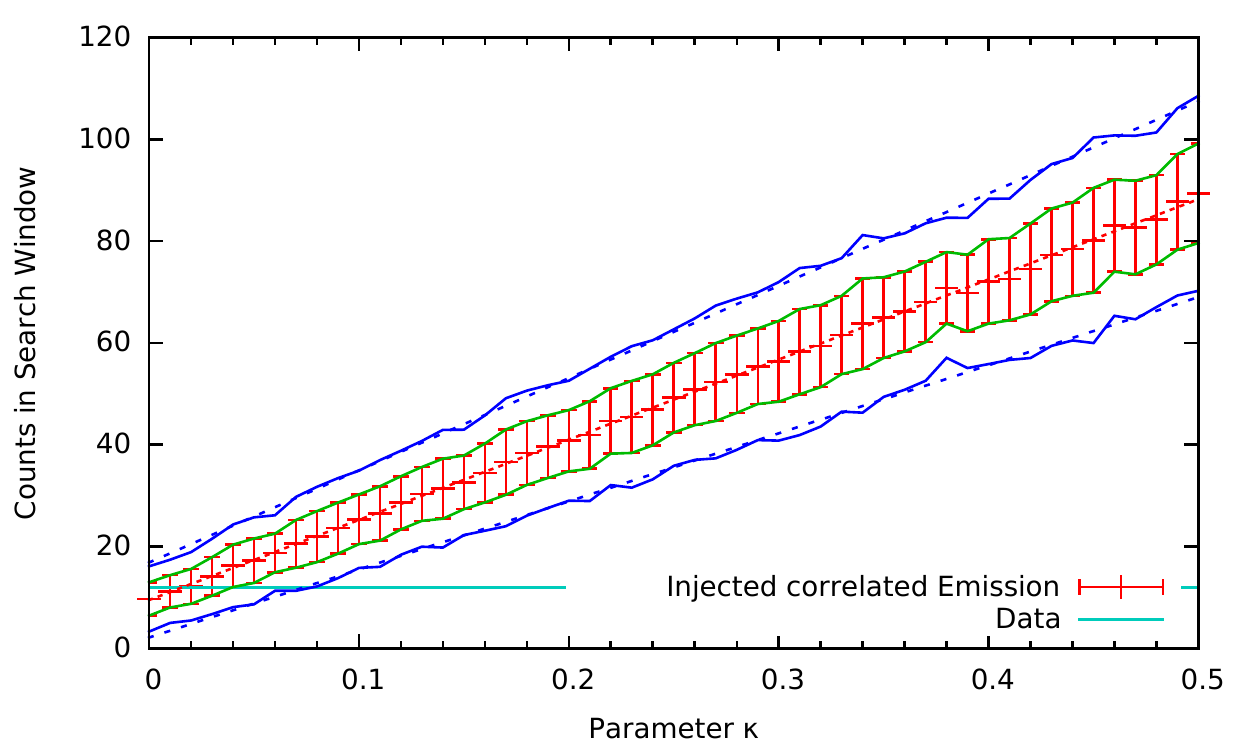}\\
   \caption{Results as described in Figure~\ref{fig:results_SW_ALL} and Figure~\ref{fig:results_SW_MP} for the P2 emission component (LFIP radio phase).}
              \label{fig:results_SW_LFIP}%
    \end{figure*}

In order to differentiate between P1 and P2 GPs (which correspond in this particular analysis to MP and LFIP as discussed in Sec.~\ref{sec:intro}), we carry out the same analysis with radio GPs only within the phase ranges 1) between $-0.02$ and 0.02 (centered on MP radio phase) and 2) between 0.37 and 0.42 (centered on LFIP radio phase, see Table~\ref{table:rotational_phases}).   
The corresponding results are shown in Figure~\ref{fig:results_SW_MP} and \ref{fig:results_SW_LFIP} and are summarized in Table~\ref{table:kbest}.

\section{Discussion}
\label{sec:discussion}
The present results do not show a statistically significant correlation between radio GPs and VHE $\gamma$-rays from the Crab pulsar. No correlation was found also in several studies carried out in the past, including the work of \cite{1974NCimB..24..153A,1995ApJ...453..433L,2011ApJ...728..110B,2012ApJ...760...64M,2012ApJ...760..136A}. 
A correlation with optical photons was found by \cite{2003Sci...301..493S}, as a 3\% higher average intensity over many periods with GPs observed. 
In order to compare this study with previous ones, it is useful to convert $\kappa$ to the factor of flux enhancement during GPs.
It can approximately be done as follows. Upper limits in number of $\gamma$-rays accompanied with a radio GPs ($N_{UL}$) are
 \begin{eqnarray}
 N_{UL} = \kappa_{UL} \cdot N_\gamma
 \end{eqnarray}
 where $N_{\gamma}$ is the number of observed pulsed $\gamma$-rays which is $443.0\pm 73.4$ as shown in Fig.~\ref{fig:phasediagram_VHE}.
Total observation time "around GPs" $T_{GP}$ can be computed from the number of obtained number of GPs $N_{GP}$ and the size of the search window $T_{SW}$ as 
 \begin{eqnarray}
 T_{GP} =  N_{GP} \cdot T_{SW} = N_{GP} \cdot P_{Crab} \cdot f_{SW}
 \end{eqnarray}
 where $f_{SW}$ is the search window in fraction of the rotation period, such as 1/9, 1/3, 1 and 3 for this study.
 Since $N_{GP}$ is 99444 as shown in Table \ref{table:summary_radio}, $T_{GP} \simeq 0.93 \cdot f_{SW}$ hours.
 
 $N_{UL} / T_{GP}$ should be compared with $N_{\gamma}/ T_{total}$, where $T_{total}$ is the total observation time which is 16 hours.
 Then, the upper limit in the flux enhancement $F_{UL}$ is written as 
 \begin{eqnarray}
 F_{UL}  &=&  (N_{UL} / T_{GP}) / (N_{\gamma}/ T_{total}) \\
  &=&  \frac{(\kappa \cdot N_\gamma) / (N_{GP} \cdot P_{Crab} \cdot f_{SW})} {N_{\gamma}/ T_{total}}\\
  &=& 17.3 \cdot \kappa / f_{SW}\label{ful}
  \end{eqnarray}
 
Therefore, the upper limit in $\kappa$ of 0.45 for $f_{SW}$ = 1 (see Table~\ref{table:kbest}) translates to 740\% of flux enhancement while $\kappa$ of 0.19 for SW = 1/9 translates to 2900\%.
This calculation shows that the sensitivity at $\gamma$-ray energies and telescope time available for this study are not sufficient to detect a statistically significant correlation or place an upper constraint comparable to the correlation observed in the optical regime. The corresponding expressions in Equation~\ref{ful} for the phase resolved analysis are F$_{UL,P1}$ = $0.848 * \kappa / f$ and F$_{UL,P2}$ = $4.04 * \kappa / f$, accounting for the number of GPs (82055 and 17041 respectively) and the shortening of T$_{total}$ due to the phase cuts.\par
The only existing theoretical prediction for a correlation at frequencies higher than 5~GHz is given by \cite{2007MNRAS.381.1190L}. However, this model is applicable to radio GPs at the phase ranges of P2 above 5~GHz, which does not cover the frequency range of the observations presented in this work, making the model not applicable to our observations. The studies by \cite{2011ApJ...728..110B} and \cite{2012ApJ...760..136A} addressed radio GPs above 5 GHz, and reported 95\% confidence level upper limits on the enhanced flux of $5-10$ times higher the flux measured by VERITAS. The higher energy threshold of VERITAS combined with the steep power-law spectrum of the Crab pulsar may have limited the sensitivity of the study. The correlation study carried out by \citet{2012ApJ...760...64M} at 300~MHz and 1.2~GHz did not result in any statistically significant findings in spite of extended searches for coincidences between radio GPs and GeV $\gamma$-rays. In the latter case  data taken by the Large Area Telescope (LAT) on board the \textit{Fermi} satellite~\citep{2009ApJS..183...46A} were used which, in comparison with the present work, provided data with 
a lower background. The data set was spanning over 15 months but the smaller collection area of the space-borne detector could be a limiting factor regarding the number of detected events which might be the reason for not detecting any correlation. The Hitomi X-ray satellite also searched for a correlation between radio GPs and soft X-rays.
\cite{2018PASJ...70...15H} report upper limits of 22\% to 80\% of the peak flux at a range of 2 to 300 keV.\\
A recent review on the radio emission physics of the Crab pulsar given by \cite{2016JPlPh..82c6302E} suggests that the observed radio and high energy emission might have origin in the same spatial regions within the magnetosphere, due to the fact that both main radio and high energy emission components appear approximately at the same phase ranges. However, a satisfactory theoretical approach still needs to be found. The variety of instabilities in the radio emission of the Crab pulsar (including GPs) leads to the assumption that the radio emission sites are dynamic and unstable \citep{2016JPlPh..82c6302E}. The connection between these regions and the high energy emission is still an open question. Since the Crab pulsar has been an object of regular monitoring campaigns at radio \citep{1993MNRAS.265.1003L} as well as at VHE $\gamma$-ray wavelengths \citep{2010A&A...523A...2M}, we suggest a coordination of the respective observations. 
Simultaneous observations at both energy ranges can lead to a further examination of the obtained results, especially below and above 5 GHz by including radio GPs from the Crab pulsar at frequencies before and after the described transition. 

\section{Summary}
\label{sec:summary}
In this work a correlation study between radio GPs and $\gamma$-rays above 60 GeV from the Crab pulsar is presented. The data used for this study were taken with the Effelsberg radio telescope (at 1347.5~MHz and 1410~MHz), the WSRT (at 1380.0~MHz) and the MAGIC telescopes (Figures~\ref{fig:phasediagram_radio}, \ref{fig:phasediagram_VHE}). The total overlap between the radio and $\gamma$-ray observations (excluding all gaps which are longer than one minute) results in 16 hours (see Table~\ref{table:summary_radio}). The approach for our correlation search is based on the idea described in \citet{2012ApJ...760..136A}, consisting of the construction of search windows around the TOA of each radio GP resulting from the radio data (see Fig.~\ref{fig:search_windows_radio}). We compare the amount of VHE $\gamma$-rays around a radio GP resulting from the observational data and MC simulations which are based on the timing characteristics of the data. To estimate the degree of correlation, we inject a variable level of a signal which is perfectly correlated with radio GPs into the simulations. With this approach we determine the fraction of VHE photons which appear to be correlated with radio GPs (indicated by component "ALL" and denoted as $\kappa_{best}$ in Table~\ref{table:kbest}).\par
Based on the described study, we conclude the following:

\begin{itemize}
    \item No statistically significant correlation between VHE pulsed photons and radio GPs at 1.4~GHz was found for search window sizes of 1/9, 1/3, 1, and 3 times the rotation period.
    \item The most stringent upper limit in the correlation degree was obtained for the search window of 1/9 of the rotational period, and not more than 19\% of the $\gamma$-rays are accompanied by GPs. This corresponds to an upper limit on the increase in pulsed flux of no more than 2900\% at 95\% confidence level.
     \item GPs in MP and LFIP are separately analyzed, and the corresponding upper limits are presented in Table~\ref{table:kbest}. Converting the correlation to a flux enhancement relative to the pulsed flux, we find upper limits between 15\% (P1 phase cut, search window of 3 P$_{Crab}$) and 85\% (P2 phase cut, search window of 1/3 P$_{Crab}$). The phase cuts do allow to place more stringent upper limits, but no statistically significant correlation could be found.
\end{itemize} 
Future observations with a larger overlap or higher sensitivity as hopefully provided by the Cherenkov Telescope Array \citep[CTA,][]{2013APh....43....3A,2019scta.book.....C} will help to provide further constraints in the still open question of a correlation between radio GPs and the VHE $\gamma$-ray emission from the Crab pulsar.

\begin{acknowledgements}
NL would like to thank Axel Jessner (MPIfR), Jean Eilek (NRAO), Maura McLaughlin (WVU), Ryan Lynch (GBO) and Vlad Kondratiev (ASTRON) for numerous helpful comments which improved the quality of the paper. We acknowledge Marina Manganaro (Croatian MAGIC Consortium) for her continuous help in preparing the manuscript. We would like to thank the anonymous referee for numerous comments which helped us to improve the quality of the manuscript. NL would also like to thank Ramesh Karuppusamy (MPIfR), Alex Kraus (MPIfR), Ralf Kisky (MPIfR), J\"org Barthel (MPIfR) and Thomas Wedel (MPIfR) for constant support during the observations with the Effelsberg radio telescope. NL gratefully acknowledges the support of this study by the ASTRON/JIVE Helena Kluyver female visitor program. 
This study is partly based on observations with the 100-m telescope of the MPIfR (Max-Planck-Institut f\"ur Radioastronomie) at Effelsberg. Another part of this study was carried out with data taken with the Westerbork Synthesis Radio Telescope (WSRT). The Westerbork Synthesis Radio Telescope is operated by the ASTRON (Netherlands Institute for Radio Astronomy) with support from the Netherlands Foundation for Scientific Research (NWO). We would like to thank the Instituto de Astrof\'isica de Canarias for the excellent working conditions at the Observatorio del Roque de los Muchachos in La Palma. The financial support of the German BMBF and MPG, the Italian INFN and INAF, the Swiss National Fund SNF, the ERDF under the Spanish MINECO (FPA2015-69818-P, FPA2012-36668, FPA2015-68378-P, FPA2015-69210-C6-2-R, FPA2015-69210-C6-4-R, FPA2015-69210-C6-6-R, AYA2015-71042-P, AYA2016-76012-C3-1-P, ESP2015-71662-C2-2-P, CSD2009-00064), and the Japanese JSPS and MEXT is gratefully acknowledged. This work was also supported by the Spanish Centro de Excelencia "Severo Ochoa" SEV-2012-0234 and SEV-2015-0548, and Unidad de Excelencia "Mar\'ia de Maeztu" MDM-2014-0369, by the Croatian Science Foundation (HrZZ) Project 09/176 and the University of Rijeka Project 13.12.1.3.02, by the DFG Collaborative Research Centers SFB823/C4 and SFB876/C3, the Polish National Research Centre grant UMO-2016/22/M/ST9/00382 and by the Brazilian MCTIC, CNPq and FAPERJ.
    \end{acknowledgements}

\bibliographystyle{aa}
\bibliography{cit.bib}
 
\appendix

\section{Description of Monte Carlo Simulations}
\label{sec:appendix_MC_sim}
The interpretation of the flux enhancements critically relies on the prediction of coincident radio and $\gamma$-ray counts from uncorrelated events. The extraction of statistical features of the present radio and $\gamma$-ray observations, as well as the construction of the simulated MC observations (which have the same statistical properties but are uncorrelated by construction), are here described step by step.
Afterwards the $\gamma$-ray events which are coincident with radio GPs in some short search window can be regarded as real, potentially correlated, observations.
\subsection{Determination of Statistical Properties - Radio Observations}
\subsubsection{Event Count}
The first property, which is matched by synthetic observation, is the total number of observed radio GPs. This number is determined for each  night of observation and is designated by N (note that there is no relation with the quantity N from Sec.\ref{sec:no-corr_sim}). 
\subsubsection{Phase Profile}
The second property reproduced is the phase bound occurrence of GPs. GPs have been observed only at the phase ranges of P1 and P2. The distribution of GPs is modeled by a Gaussian function. Since the observational GP data used for this study only includes single pulses brighter than 5 times the rms (7 times the rms in the case of Effelsberg data) of the raw data, they contain few pulses outside of the average emission components. One set of statistical parameters for each observing night are the amplitude a, the phase m and the width s for both regular emission components P1 and P2. The data are then modeled by the probability p that a GP arrives at a rotational phase $\varphi$.
\begin{eqnarray}
p(\varphi) = \frac{a_1}{\sqrt{2\pi}s_1}\exp{(-\frac{1}{2} \frac{(\varphi - m_1)^2}{s_1^2})} + \frac{a_2}{\sqrt{2\pi}s_2}\exp{(-\frac{1}{2} \frac{(\varphi - m_2)^2}{s_2^2})}
\end{eqnarray}

The normalization of p is such that the integral over all probabilities $\int^1_0 p(\varphi){\rm d}\varphi = N$, where $N$ is the total number of GPs observed in the respective night. The rotational phase $\varphi$  is restricted to the range 0 . . . 1 and the probability density is aliased to this range. Hence any remaining nonzero probability for the arrival at $\varphi$ = 1.01 is added to the probability for an arrival at rotational phase $\varphi$ = 0.01 (of the next rotation). The fitting procedure which is used to determine values of $a$, $m$ and $s$ from the observed data operates on binned rotational phases $\varphi$ with 1000 bins for the interval 0 . . . 1. The procedure is first done jointly for all observations that were carried out with one telescope at one frequency (see Table\ref{table:rotational_phases} for details). The results of this first fit, made robust by the large number of available events, are then used as starting parameters for the individual fitting of all observation nights.
 \subsubsection{Interarrival Times}
The third set of statistical properties represented is the distribution of
interarrival times between successive GPs. The interarrival times are modeled directly from observed separations. To be able to do this, the interval between successive GPs was calculated and stored in a list. Excessively large intervals, e.g. above 30 seconds, were discarded. The usage of the list of interarrival times in the Markov Chain Monte Carlo (MCMC) simulation of synthetic observations is described below.
 \subsection{Determination of Statistical Properties - $\gamma$-ray Observations}
\subsubsection{Phase Profile}
The $\gamma$-ray observations employed in the present work have a substantially higher level of background emission and therefore need to be modeled differently. The distribution of $\gamma$-ray events over rotational phase is very similar to the respective distribution for radio GPs, but includes a constant background a3:
 \begin{eqnarray}
p(\varphi) = \frac{a_1}{\sqrt{2\pi}s_1}\exp{(-\frac{1}{2} \frac{(\varphi - m_1)^2}{s_1^2})} + \frac{a_2}{\sqrt{2\pi}s_2}\exp{(-\frac{1}{2} \frac{(\varphi - m_2)^2}{s_2^2})} +a3
\end{eqnarray}

To determine all parameters in this equation, all observed $\gamma$-ray events are binned into 200 phase bins and fitted with the above equation.
\subsubsection{Event Counts}
We refer to the fact that the total number of events N is split into a number of "background" events
\begin{eqnarray}
N_{\rm off} = \int^1_0 a3 {\rm d} \varphi
\end{eqnarray}
and "pulsed" events
\begin{eqnarray}
N_{\rm 1,2} = \int \frac{a_{1,2}}{\sqrt{2\pi} s_{1,2}} \exp{(-\frac{1}{2} \frac{(\varphi - m_{1,2})^2}{s_{1,2}^2})}  {\rm d}\varphi
\end{eqnarray}

 \subsubsection{Trigger Rates}
The distribution of interarrival times in this case is dominated by background events and is therefore not as relevant as for the radio data. However, there is another effect which needs to be modeled in order to avoid systematic differences between real and synthetic observations generated from the MCMC simulation: the telescope performance varies according to zenith angles, moonlight and weather conditions. To take that in account, the raw trigger rate is determined for every one second interval and is later used to create events which follow the correct arrival rate over the course of the night.
The raw trigger rate is used because of its stability respect to discrete Poisson noise. 
While examining the data, we found that the fraction of $\gamma$-ray events is constant even during changes in the trigger rate.

\subsubsection{Identification of Time Windows with Multi-wavelength Coverage}
The observations at radio wavelengths and $\gamma$-rays show gaps in coverage due to unfavorable weather conditions, switches between sub-runs during data-taking and hardware problems. Consequently, only time windows during which observations at both wavelength ranges were successfully carried out can be used for our data analysis. The existence of gaps in the middle of the observing night might alter the overall statistics and therefore has to be modeled in the MCMC simulations as well.
We assume that a gap exists in the multi-wavelength coverage when no radio GP or $\gamma$-ray event has been received for one minute. The currently open observing window is closed at the arrival time of the last event before the gap and a new window is tentatively started. If the length of a resulting window is below the cutoff length, it is omitted. 
\subsection{Determination of the Arrival Rate of (nearly) Coincident Events}
The fundamental idea of the analysis method is to take a radio GP, take a small time window around it and to count the number of $\gamma$-ray events in that window. This is repeated for all radio GPs and the total number of $\gamma$-ray events in all such search windows is added up. At the end the total number is divided by the number of radio GPs and the duration of the search window to give an estimate of the $\gamma$-ray count rate which is nearly coincident with radio GPs for the given width of the search
window.
Emission processes that lead to the production of both, radio GPs and $\gamma$-rays, but with some small time lag between the two, cannot be excluded. Therefore the analysis process does not only use one SW, centered around the arrival time of the radio GP, but two additional SWs of the same length. The "before" window covers the time span before the start of the centered window and the "after" window starts at the end of the centered window. See Fig.~\ref{fig:search_windows_radio} for an illustration of the relative timing of the search windows.
Other staggerings are imaginable, such as an "early" window which covers a time interval exactly up to the BAT (Barycentered Arrival Time) of the radio GP, or a "late" window that starts at this BAT. However, such staggering does not allow for the following implementation method that is used to efficiently compute the arrival rates for search windows of different length. The shortest SWs cover 1/9 of one pulsar rotation and there are separate "before", "centered" and "after" windows. The three SWs together are 1/3 of a pulsar rotation and are centered around the arrival time of the radio GP. In other words, they cover exactly the "centered" SW for the duration of 1/3 rotation. Together with the "before" and "after" windows of 1/3 rotation each, one pulsar rotation is covered. This nested construction continues all the way the "centered" window of the longest duration, 37 rotations.
In total the 10 different SWs durations require 21 nested windows, which can be computed during a single pass through the data. To output the arrival rates for a given duration, the windows of longer durations are ignored and all windows of shorter durations are summed to determine the number of events in the "centered" window. This method has reduced the analysis run time by nearly one order of magnitude, compared to an initial, more naive implementation that computed the arrival rates for each duration of the SW separately. Further implementation methods are required to find all $\gamma$-ray events that fall into the search window of a radio GP with acceptable performance. First of all, all events are loaded into memory, checked against the time windows of overlapping multi-wavelength coverage and stored in two lists for radio and $\gamma$-ray BATs respectively. Both lists are afterwards sorted in time. The analysis script then loops over all radio GPs in the first list. The monotonously increasing BAT of the radio GPs and the sorted nature of the $\gamma$-ray BAT list allows two important optimization steps: 1) Every $\gamma$-ray event that happened before the start of the "before" window with the longest duration relative to the last radio event, comes too early in time to fall into any search window relative to the current radio GP. 2) Once one $\gamma$-ray event falls past the end of the "after" search window of the longest duration, no subsequent $\gamma$-ray event will fall into any search window relative to the current radio GP.
This procedure reduced the number of $\gamma$-ray events which need to be considered significantly. The calculation of the time difference between the $\gamma$-ray event and the radio GP and the conversion into pulsar periods can be skipped entirely for a large fraction of $\gamma$-ray events. The computation is not particularly time-consuming, but the large number of combinations between every radio event and every $\gamma$-ray
event would produce a large overall run time. The first optimization requires that a marker is maintained for the last $\gamma$-ray event which was too early for the previous radio GP. When the iteration for the current radio GP starts at that point in the list of $\gamma$-ray BATs and an event is found which is before the start of the first search window, the marker needs to be incremented accordingly. This state is very easy and cheap to maintain and incurs no noticeable performance cost.

\end{document}